\documentclass[12pt,journal,draftclsnofoot,onecolumn]{IEEEtran} 

\usepackage{amssymb}
\usepackage{amsmath}
\usepackage{amsmath,bm}
\usepackage{amsthm}
\usepackage{graphicx}
\usepackage{subfigure}
\usepackage{multirow}
\usepackage{makecell} 
\usepackage{cite}
\usepackage{enumerate}
\usepackage{enumitem}
\usepackage[ruled]{algorithm2e}
\usepackage{algorithmic}  
\usepackage{color, soul}
\usepackage[labelsep=period]{caption}
\allowdisplaybreaks[4]
\SetKw{In}{Initialize}
\SetKw{Input}{Input}
\SetKw{Output}{Output}

\newenvironment{indexterms}{
	\begin{center}\small\bfseries Index Terms\end{center}
	\begin{quote}\small
	}
	{\end{quote}}

\begin{document}
	\newtheorem{theorem}{\bf~~Theorem}
	\newtheorem{remark}{\bf~~Remark}
	\newtheorem{observation}{\bf~~Observation}
	\newtheorem{definition}{\bf~~Definition}
	\newtheorem{lemma}{\bf~~Lemma}
	\newtheorem{preliminary}{\bf~~Preliminary}
	\newtheorem{proposition}{\bf~~Proposition}
	\renewcommand\arraystretch{0.9}
	
	 \title{\setlength{\baselineskip}{3pt}\Large Intelligent Omni-Surfaces: Reflection-Refraction Circuit Model, Full-Dimensional Beamforming, and System Implementation}
	\author{\normalsize \IEEEauthorblockN{
			{Shuhao Zeng}, \IEEEmembership{\normalsize Student Member, IEEE},
			{Hongliang Zhang}, \IEEEmembership{\normalsize Member, IEEE},\\
			{Boya Di}, \IEEEmembership{\normalsize Member, IEEE},
			{Yuanwei Liu}, \IEEEmembership{\normalsize Senior Member, IEEE},\\
			{Marco Di Renzo}, \IEEEmembership{\normalsize Fellow, IEEE},
			{Zhu Han}, \IEEEmembership{\normalsize Fellow, IEEE},\\
			{H. Vincent Poor}, \IEEEmembership{\normalsize Life Fellow, IEEE},
			{and Lingyang Song}, \IEEEmembership{\normalsize Fellow, IEEE}}
		\thanks{S. Zeng, B. Di, and L. Song are with School of Electronics, Peking University, Beijing 100871, China (email: shuhao.zeng@pku.edu.cn; boya.di@pku.edu.cn; lingyang.song@pku.edu.cn).}
		\thanks{H. Zhang and H. V. Poor are with Department of Electrical and Computer Engineering, Princeton University, Princeton, NJ 08544, USA (email: hz16@princeton.edu, poor@princeton.edu).}
		\thanks{Y. Liu is with the School of Electronic Engineering and Computer Science, Queen Mary University of London, London E1 4NS, U.K. (e-mail: yuanwei.liu@qmul.ac.uk).}
		\thanks{M. Di Renzo is with Universit\'{e} Paris-Saclay, CNRS, CentraleSup\'{e}lec, Laboratoire des Signaux et Syst\`{e}mes, 3 Rue Joliot-Curie, 91192 Gif-sur-Yvette, France. (marco.di-renzo@universite-paris-saclay.fr)}
		\thanks{Z. Han is with Electrical and Computer Engineering Department, University of Houston, Houston, TX, USA, and also with the Department of Computer Science and Engineering, Kyung Hee University, Seoul, South Korea (email: zhan2@uh.edu).}
	}
	
	\maketitle
	\vspace{-2cm}
	\begin{abstract}
	\vspace{-0.3cm}
	The intelligent omni-surface~(IOS) is a dynamic metasurface that has recently been proposed to achieve full-dimensional communications by realizing the dual function of anomalous reflection and anomalous refraction. Existing research works provide only simplified models for the reflection and refraction responses of the IOS, which do not explicitly depend on the physical structure of the IOS and the angle of incidence of the electromagnetic~(EM) wave. Therefore, the available reflection-refraction models are insufficient to characterize the performance of full-dimensional communications. In this paper, we propose a complete and detailed circuit-based reflection-refraction model for the IOS, which is formulated in terms of the physical structure and equivalent circuits of the IOS elements, as well as we validate it against full-wave EM simulations. {Based on the proposed circuit-based model for the IOS, we analyze the asymmetry between the reflection and transmission coefficients.} Moreover, the proposed circuit-based model is utilized for optimizing the hybrid beamforming of IOS-assisted networks and hence improving the system performance. To verify the circuit-based model, the theoretical findings, and to evaluate the performance of full-dimensional beamforming, we implement a prototype of IOS and deploy an IOS-assisted wireless communication testbed to experimentally measure the beam patterns and to quantify the achievable rate. The obtained experimental results validate the theoretical findings and the accuracy of the proposed circuit-based reflection-refraction model for IOSs.
	
	

	\end{abstract}
\vspace{-0.3cm}
	\begin{indexterms}
		\vspace{-0.4cm}
	Intelligent omni-surface, circuit-based reflection-refraction model, full-dimensional beamforming, prototype.
	\end{indexterms}
\vspace{-.8cm}
	\section{Introduction}
	The enormous increase of mobile devices and applications in the past decade has triggered the needs for new communication paradigms and technologies~\cite{E_2011}. Notably, to support high-speed and seamless data services in future wireless systems, a variety of transmission techniques that exploit the implicit randomness of the wireless environment have received increasing attention, such as spatial modulation~\cite{MHASL_2014} and massive multiple input and multiple output~(MIMO) systems~\cite{EOFT_2014}. However, multiple-antenna technologies usually require a high hardware cost and power consumption, and need complex signal processing algorithms~\cite{M_2020,Y_2021,QMM}. Existing techniques, in addition, only adapt themselves to the wireless propagation environment. Therefore, the quality of service cannot always be guaranteed in harsh propagation environments, e.g., when the line-of-sight links are blocked at high frequency bands~\cite{G_arxiv,BMM_arxiv}. 
	
	Thanks to the recent developments and research advances in the field of metasurfaces, a new wireless technology named \emph{reconfigurable intelligent surface}~(RIS) is being developed, which is intended to turn the wireless propagation environment into a programmable entity for further improving the performance of state-of-the-art technologies in challenging propagation environments~\cite{HBLZ_2021,THSMB_arxiv}. Specifically, an RIS is an ultra-thin surface consisting of multiple subwavelength scattering elements~\cite{MMDAMCVGJHJAGM_2019}, which can apply an adjustable amplitude and phase shift to the incident electromagnetic~(EM) waves~\cite{BHLYZH-2020,MFS_arxiv}. A typical implementation of an RIS comprises several scattering elements with integrated electronic circuits, such as  positive-intrinsic-negative~(PIN) diodes or varactors~\cite{L_2020,RPDAMJ}. By controlling the ON/OFF status of the PIN diodes or the biasing voltages of the varactors, the amplitude and the phase of the incident EM waves can be adjusted~\cite{SHBZL_2021,MHLKZG_2020}. {Benefiting from this programmable feature, an RIS can shape the propagation environment into a desirable form, and can hence improve the link quality and the coverage even in harsh and challenging propagation environments~\cite{BMM_arxiv,THSMB_arxiv}.} In addition, an RIS has the appealing attribute of shaping the wireless environments without requiring power amplifiers, digital processing units for signal regeneration, and multiple radio frequency chains. Even though some power is still needed to make the surface reconfigurable~\cite{WMXJYMYSQT}, these characteristics may reduce the total power consumption and the hardware cost as compared with other technologies. Some examples are available in~\cite{MCDD_2020,KMJOK_2021,HBKZHL_2022}. These features collaboratively make an RIS a promising technology for future wireless networks, but they result in some design and deployment challenges, which include how to efficiently estimate the channel and how to optimize the RISs in a scalable manner, see, e.g.,~\cite{Y_2021,HBLZ_2021}.

	The vast majority of research works conducted in the field of RIS-aided communications have considered RISs that operate in reflection mode. In other words, the RIS is designed to reflect the signals that impinge on one side of the surface towards user equipments~(UEs) that are located on the same side, while the transmitted signal is minimized~\cite{BHLYZH-2020}. This implementation is also referred to as intelligent reflecting surface~(IRS). Because of the reflection characteristics of the IRS, however, the UEs located on the opposite side of the surface are out of coverage. 
	Use cases where UEs located on the side of the surface that is not illuminated by the transmitter may benefit from the deployment of an RIS have recently been reported in~\cite{KMJOK_2021}. {Motivated by these considerations, an RIS design that is referred to as intelligent omni-surface~(IOS) has recently been introduced~\cite{SHBYMZHL}. In the literature, this design of RIS is also referred to as simultaneously transmitting and reflecting RIS~(STAR-RIS)~\cite{Y_online}.} In contrast to an IRS, an IOS has the dual functionality of reflection and refraction, which makes it capable of providing ubiquitous coverage to the UEs located on both sides of the surface, so as to achieve full-dimensional wireless communications. 
	
	The existing works on IOSs are limited in number and are focused on physical and channel modeling~\cite{JYXJLHL_arxiv,Y_online}, beamforming design~\cite{SHBYMZHL,HSBYMDLZH,YBHZHL_2022,JYXO_2021}, and performance analysis~\cite{SHBYZHL_2021,CWYZL_arxiv}. Specifically, several IOS hardware and channel models as well as the available hardware implementations are overviewed in~\cite{JYXJLHL_arxiv}, which serve as a physics-compliant pipeline for further investigations. {In~\cite{Y_online}, signal models and  operation protocols for IOSs are investigated, based on which several promising application scenarios are introduced and discussed.} To maximize the sum rate of an IOS-aided communication system, the authors of~\cite{SHBYMZHL} jointly optimize the IOS configuration and the digital beamformer at the base station~(BS). The proposed hybrid beamforming scheme is further validated with the aid of a hardware testbed in~\cite{HSBYMDLZH}. The authors in~\cite{YBHZHL_2022} then extend to a multi-cell IOS-aided network,  where a distributed rather than centralized hybrid beamforming scheme is designed to maximize the sum rate so that no exchange of channel state information is required between the cells. The authors of~\cite{JYXO_2021} investigate the performance of IOS-aided non-orthogonal multiple access~(NOMA) and IOS-aided orthogonal multiple access~(OMA) networks, by jointly designing the IOS configuration and the resource allocation scheme. Motivated by the multiplicity of signal transformations that an RIS can apply, i.e., reflections~(IRS), refractions, and joint reflections and refractions~(IOS), the authors of~\cite{SHBYZHL_2021} consider an IOS-aided downlink multi-user communication network, and the corresponding system capacity is compared against that offered by surfaces that can either only reflect or refract the incident signals. In~\cite{CWYZL_arxiv}, the diversity gain and outage probability of an IOS-aided downlink NOMA network with randomly distributed UEs are analyzed. 
	
	The reflection-refraction models for IOSs that are employed in the existing works are, however, oversimplified, which poses some limitations on the case studies that can be analyzed and on the conclusions that can be drawn from them. Specifically, the relationship between the physical implementation of the IOS and the reflection and transmission coefficients is not explicitly given in existing research works. Therefore, the components of the IOS that can be optimized to obtain the desired reflection and refraction properties are uncertain. Besides, the existing models do not embody the impact of the angle of incidence on the reflection and transmission coefficients. {As a result, the reflection and transmission coefficients employed to design the hybrid beamforming can be inaccurate, leading to system performance degradation.}

	
	
	
	The development of a circuit-based and electromagnetically consistent reflection-refraction model that embodies the aforementioned properties is, however, challenging, since the reflection and transmission coefficients are coupled. Therefore, it is challenging to construct a unified equivalent circuit-based model that accurately describes the reflection and transmission coefficients simultaneously. In this paper, we introduce a new circuit-based reflection-refraction model for a generic IOS element based on circuits theory. {Based on the proposed model, in addition, we analyze the asymmetry between the reflection and transmission coefficients, and introduce a hybrid beamforming scheme to enhance the system performance.} To verify the circuit-based model and to evaluate the performance of  full-dimensional beamforming, we implement a prototype of IOS and deploy an IOS-assisted wireless communication testbed. {Experimental measurements are conducted to quantify the performance of full-dimensional communications in terms of beam patterns and data rate.} Specifically, the contributions of this paper are as follows.
	\begin{itemize}
		\item We introduce a circuit-based model for the reflection and transmission coefficients of the IOS elements. The proposed model  explicitly depends on the physical structure of the IOS and on the angle of incidence of the EM waves. The EM consistency and the accuracy of the proposed model are, in addition, validated with the aid of full-wave EM simulations.
		
		\item Based on the obtained model, we analyze the asymmetry between the reflection and transmission coefficients, i.e., the reflection and transmission coefficients are shown to be different in terms of amplitude and phase responses. Besides, the proposed model is utilized for optimizing the hybrid beamforming of IOS-assisted networks and thus improving the system performance. 
		
		\item We design and implement an IOS, and we deploy an IOS-assisted wireless communication testbed to experimentally measure the beam patterns, to quantify the achievable rate, and to verify the accuracy of the circuit-based model.

%

		
		
	\end{itemize}
	
	The rest of this paper is organized as follows. In Section~\ref{model}, we introduce the concept of IOS, we propose a new circuit-based model for the reflection and transmission coefficients for the IOS elements, and we validate the EM consistency of the model through full-wave EM simulations. In Section~\ref{bf}, we introduce a hybrid beamforming scheme based on the proposed circuit-based model. In Section~\ref{prototype}, we report an IOS hardware prototype, and we illustrate a testbed of an IOS-assisted wireless communication system. In Section~\ref{metrics_process} and Section~\ref{verify_all}, we report the adopted method for performing the measurements and the obtained experimental results, respectively. Finally, conclusions are drawn in Section~\ref{conc}.
	\section{Circuit-Based Reflection and Refraction Model for IOSs}
	\vspace{-.2cm}
	This section is organized as follows. The main characteristics of an IOS are introduced in Section~\ref{basic}; a new circuit-based reflection and refraction model is proposed in Section~\ref{explicit_model} and is then validated with the aid of full-wave simulations in Section~\ref{verify}.
	\vspace{-0.4cm}
	\label{model}
	\subsection{Basics of Intelligent Omni-Surfaces}
	\label{basic}
	\vspace{-0.2cm}
\begin{figure*}[!t]
	\centering
	\subfigure[Structure of an IOS element]{
		\begin{minipage}[b]{0.45\textwidth}
			\centering
			\includegraphics[width=0.8\textwidth]{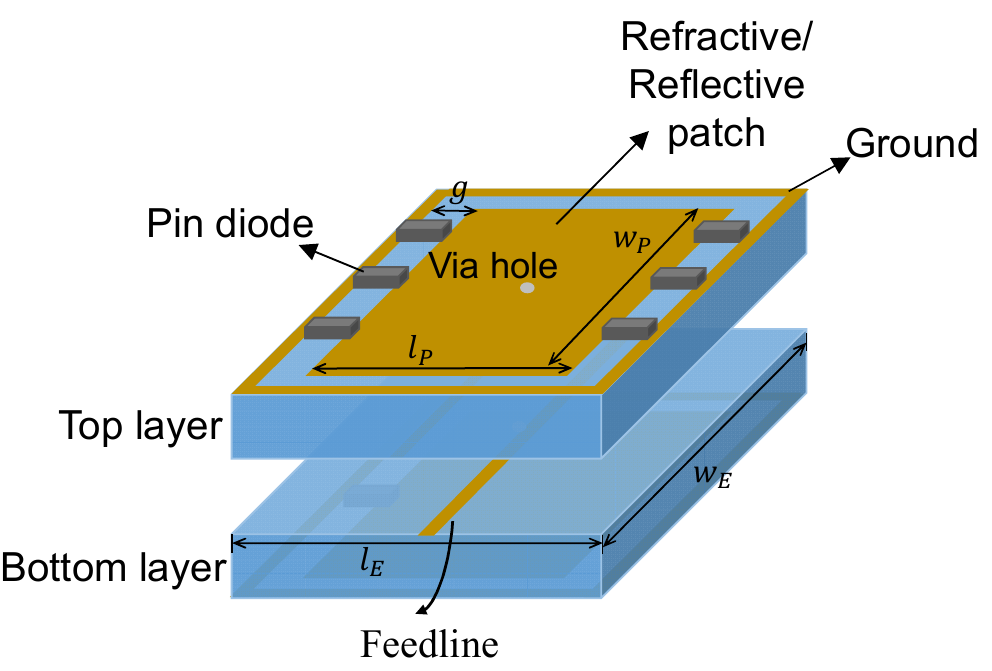}
			\vspace{-0.2cm}
			\label{angle_stru}
	\end{minipage}}
	\subfigure[Photo of an implemented IOS element]{
		\begin{minipage}[b]{0.45\textwidth}
			\centering
			\includegraphics[width=0.8\textwidth]{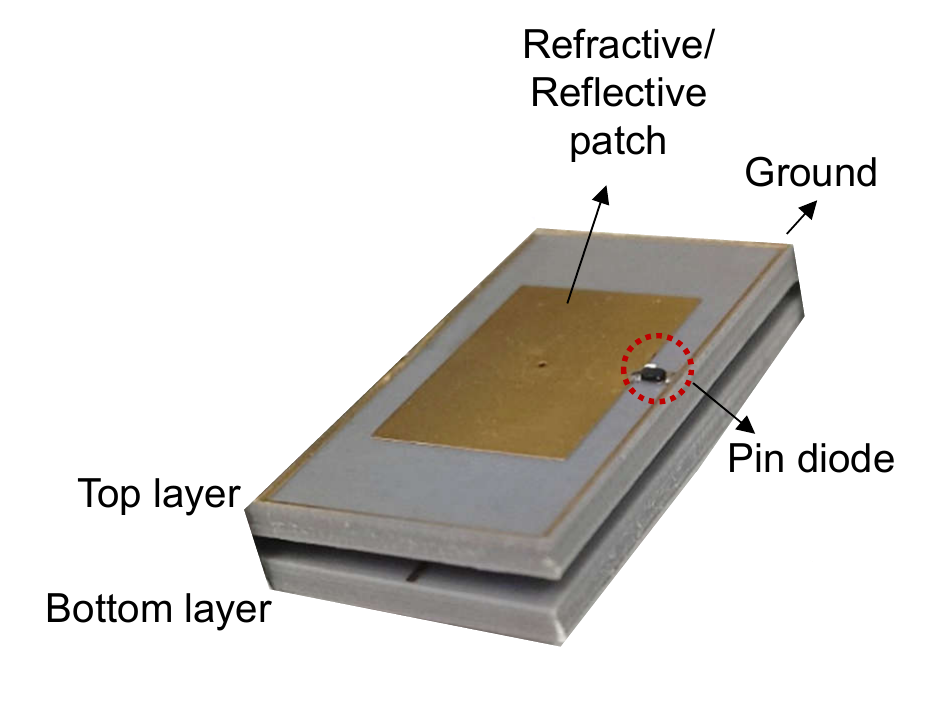}
			\vspace{-0.2cm}
			\label{angle_imp}
	\end{minipage}}
	\vspace{-4mm}
	\caption{Structure and one example of an IOS element.}
	\label{angle}
	\vspace{-9mm}
\end{figure*}

	An IOS is a two-dimensional array of electrically controllable scattering elements with equal size. Each reconfigurable element consists of two symmetrical layers, and each layer contains one metallic patch and $N$ PIN diodes that are evenly distributed on a dielectric substrate, as shown in Fig.~\ref{angle}. The metallic patch is connected to the ground via the PIN diodes. According to predetermined bias voltages, the PIN diodes can be switched between their ON and OFF states, and the state of each IOS element is determined by the states of the PIN diodes on the element. At the bottom of each layer, there is a feedline that is connected to the metallic patch through a via hole. The feedline is utilized to provide the required bias voltages to the PIN diodes. 
	
	When an EM wave impinges upon the IOS element, it excites time-varying currents within the element, which reradiate the reflected and refracted EM waves. When the IOS element is configured to a different state, the surface impedance of the element changes accordingly, which in turn has an impact on the excited currents and the reradiated EM waves~\cite{JYXO_2021_STAR}. Therefore, the reflection and refraction properties of the IOS element depend on the state of the element, i.e., the configuration of the PIN diodes. Among all the possible states of the IOS element, $N_S$ states are selected in order to control the EM waves. The set of selected states is denoted by $\mathcal{S}$. To completely characterize the reflective and refractive characteristics of the IOS, a model for the reflection and transmission coefficients is required. This is elaborated in the next subsection.
	\vspace{-0.5cm}
	\subsection{Circuit-based Reflection-Refraction Model}
	\label{explicit_model}
\begin{figure}[!t]
		\centering
		\includegraphics[width=0.9\textwidth]{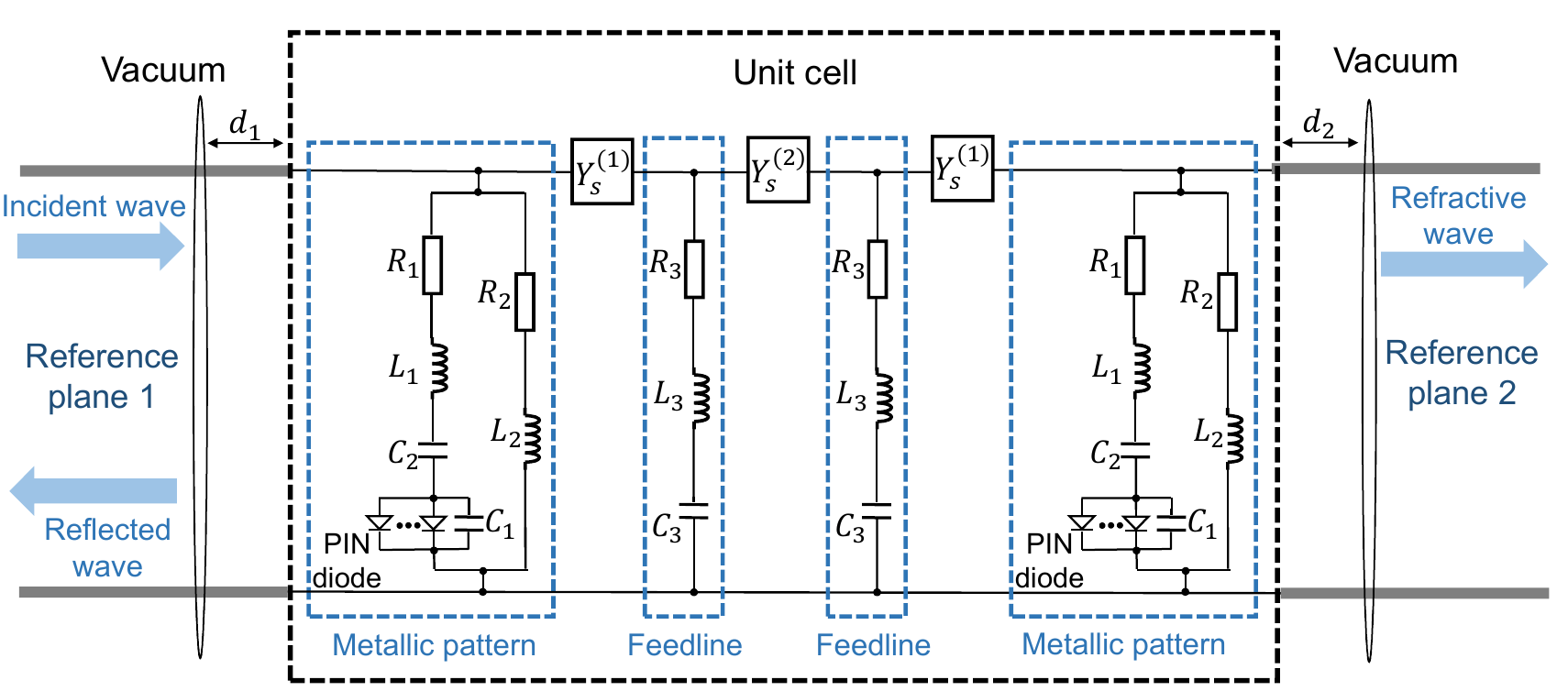}
		\vspace{-2mm}
		\caption{Equivalent circuit model of the IOS element.}
		\label{equivalent_circuit}
		\vspace{-11mm}
	\end{figure}

	To model the reflection and refraction properties of an IOS element\footnote{The main objective of this paper is to characterize the impact of the IOS physical structure and the angle of incidence on the reflection and transmission coefficients. Therefore, we do not consider the effect of  polarization, and assume that the incident signal is a transverse electric~(TE) polarized wave as a case study.}, we introduce an equivalent circuit model, as shown in Fig.~\ref{equivalent_circuit}. The proposed model has the following characteristics. The vacuum on both sides of the IOS is modeled by two semi-infinite transmission lines~\cite{VFMRMF_2016} and the IOS element is modeled as a two-port microwave network. The characteristic impedance $Z_0$ and the propagation constant $\beta$ of the equivalent transmission line are modeled by the impedance of free space and the wave number in vacuum, respectively, i.e., $Z_0=377~\Omega$ and $\beta=\frac{2\pi}{\lambda}$~\cite{K_2007}, where $\lambda$ is the wavelength of the EM wave in vacuum. Since the IOS element contains four metallic layers, i.e., the metallic pattern on the top layer, the feedline on the top layer, the feedline on the bottom layer, and the metallic pattern on the bottom layer, the equivalent two-port network model consists of four metallic layers and three coupling admittances between these layers. In the following, we  elaborate on these metallic layers and coupling admittances.
	
	As illustrated in Fig.~\ref{equivalent_circuit}, each metallic layer of the IOS element is modeled with an equivalent parallel admittance that is represented by an RLC circuit. The RLC equivalent circuit for the metallic pattern consists of two capacitors, two inductors, one resistor, and several PIN diodes. Specifically, $C_1$ and $C_2$ describe the capacitances formed by the patch and the ground. Both $C_1$ and $C_2$ are positively correlated with the width $w_E$ of the IOS element and with the width $w_P$ of the metallic patch, and they are negatively correlated with the gap $g$ between the patch and the ground~\cite{F_2014}. The parameters $L_1$ and $R_1$ represent the inductance and the resistance of the central metallic patch, respectively, where the inductance $L_1$ increases with the length $l_P$ of the patch~\cite{F_2014}. Moreover, $L_2$ and $R_2$ model the inductance and resistance generated by the ground, respectively, where $L_2$ is positively correlated with the length $l_E$ of the ground. Each PIN diode can be modeled with an equivalent RLC circuit as well, whose structure depends on the state of the PIN diode~\cite{L_2019}. The RLC equivalent circuit of the feedline is constituted by the series of the resistance $R_3$, the inductance $L_3$, and the capacitance $C_3$, where $R_3$ and $L_3$ are generated by the feedline while the capacitance $C_3$ is formed by the feedline of the IOS element and the feedline of the neighbouring element. Specifically, the capacitance $C_3$ increases with the length $w_E$ of the feedline and decreases with the length $l_E$ of the element. Also, the larger the width $w_F$ of the feedline, the larger the inductance $L_3$. {The coupling effects between adjacent metallic layers can be described by coupling admittances. Since the IOS consists of two symmetric layers, the coupling admittance between the metallic pattern and the feedline on the top layer is the same as that between the metallic pattern and the feedline on the bottom layer. These admittances are represented by $Y^{(1)}_S$. Besides, $Y^{(2)}_S$ is the coupling admittance between the feedline on the top layer and the feedline on the bottom layer.}
	
	Based on the equivalent two-port network illustrated in Fig.~\ref{equivalent_circuit}, the ABCD transmission matrix of the obtained network can be expressed as~\cite{D_2005}
	{\setlength\abovedisplayskip{0cm}
		\setlength\belowdisplayskip{0.3cm}
\begin{align}
\label{ABCD_matrix}
\left[
\begin{matrix}
A & B\\
C & D
\end{matrix}
\right]\!=\!
\left[
\begin{matrix}
1 & 0\\
Y_P^{U,M} & 1
\end{matrix}
\right]
\left[
\begin{matrix}
1 & \frac{1}{Y_s^{(1)}}\\
0 & 1
\end{matrix}
\right]
\left[
\begin{matrix}
1 & 0\\
Y_P^{U,F}& 1
\end{matrix}
\right]
\left[
\begin{matrix}
1 & \frac{1}{Y_s^{(2)}}\\
0 & 1
\end{matrix}
\right]
\left[
\begin{matrix}
1 & 0\\
Y_P^{L,F}& 1
\end{matrix}
\right]
\left[
\begin{matrix}
1 & \frac{1}{Y_s^{(1)}}\\
0 & 1
\end{matrix}
\right]
\left[
\begin{matrix}
1 & 0\\
Y_P^{L,M} & 1
\end{matrix}
\right],
\end{align}
	where $Y_P^{U,M}$, $Y_P^{U,F}$, $Y_P^{L,F}$, and $Y_P^{L,M}$ denote the parallel admittances that correspond to the metallic pattern on the top layer, the feedline on the top layer, the feedline on the bottom layer, and the metallic pattern on the bottom layer, respectively.} These admittances can be expressed as
	{\setlength\abovedisplayskip{0cm}
		\setlength\belowdisplayskip{0cm}
\begin{align}
Y_P^{U,M}=\bigg(R_1+j\omega L_1+(j\omega C_2)^{-1}+\Big(\sum_{i\in \mathcal{U}}Y_{pin}^i+j\omega C_1\Big)^{-1}\bigg)^{-1}+(R_2+j\omega L_2)^{-1},
\end{align}
\begin{align}
Y_P^{U,F}=Y_P^{L,F}=\left(R_3+j\omega L_3+(j\omega C_3)^{-1}\right)^{-1},
	\end{align}
\begin{align}
Y_P^{L,M}=\bigg(R_1+j\omega L_1+(j\omega C_2)^{-1}+\Big(\sum_{i\in \mathcal{L}}Y_{pin}^i+j\omega C_1\Big)^{-1}\bigg)^{-1}+(R_2+j\omega L_2)^{-1},
\end{align}
	respectively, where $\omega$ is the angular frequency, $Y_{pin}^i$ denotes the admittance of the $i$-th PIN diode, which is determined by the state of the PIN diode, and $\mathcal{U}$ and $\mathcal{L}$ represent the set of PIN diodes on the upper and lower metallic layers of the IOS element, respectively.}
	
	Based on the ABCD transmission matrix in (\ref{ABCD_matrix}), the reflection and transmission coefficients of the considered IOS element can be formulated as~[31, Table 4.2]
	{\setlength\abovedisplayskip{0.5cm}
		\setlength\belowdisplayskip{0cm}
	\begin{align}
		\label{reflection}
		\Gamma_r=\frac{(A+B/Z_0)-Z_0(C+D/Z_0)}{(A+B/Z_0)+Z_0(C+D/Z_0)}\exp\left(-j2\beta d_1\right),
	\end{align}
	and}
{\setlength\abovedisplayskip{0cm}
	\setlength\belowdisplayskip{0.5cm}
	\begin{align}
		\label{transmission}
		\Gamma_t=\frac{2}{(A+B/Z_0)+Z_0(C+D/Z_0)}\exp\left(-j\beta (d_1+d_2)\right),
	\end{align}
	respectively, where $d_1$ and $d_2$ represent the distances between the reference planes and the surface of the IOS element.} The terms $\exp\left(-j2\beta d_1\right)$ in (\ref{reflection}) and $\exp\left(-j\beta (d_1+d_2)\right)$ in (\ref{transmission}) account for the distances with respect to the IOS at which the reflection and transmission coefficients are measured, e.g., when analyzing the IOS element with the aid of full-wave EM simulations. Therefore, $d_1$ and $d_2$ are fixed throughout this paper, and they cannot be designed to optimize the reflection and transmission coefficients. The reflection-refraction model for the IOS proposed in (\ref{reflection}) and (\ref{transmission}) depends on the specific circuital implementation of the IOS element and the tuning circuits, i.e., the PIN diodes. Therefore, it offers a realistic and accurate model for wireless applications. As far as (\ref{ABCD_matrix}), (\ref{reflection}), and (\ref{transmission}) are concerned, three remarks are in order.
	\vspace{-0.3cm}
	\begin{remark}
		The IOS cannot be configured to operate as a purely reflecting surface, i.e., we always have $|\Gamma_t|\neq0$. If, however, we ensure that $|(A+B/Z_0)+Z_0(C+D/Z_0)|$ is sufficiently large, the refracted power can be made small.
		
		On the contrary, a purely refracting IOS element is obtained by setting $(A+B/Z_0)=Z_0(C+D/Z_0)$. In this case, the reflection and transmission coefficients in (\ref{reflection}) and (\ref{transmission}) reduce to $\Gamma_r=0$ and $\Gamma_t=(A+B/Z_0)^{-1}\exp\left(-j\beta (d_1+d_2)\right)$, respectively.
	\end{remark}
\vspace{-0.7cm}
	\begin{remark}
		It is worth noting that the ABCD transmission matrix in (\ref{ABCD_matrix}), and the reflection and transmission coefficients in (\ref{reflection}) and (\ref{transmission}), respectively, are derived under the so-called \textbf{locally periodic boundary condition}~\cite{MFS_arxiv}. In other words, we consider an IOS element whose equivalent circuit is given in Fig.~\ref{equivalent_circuit}, in which the PIN diodes are set to a given state. Then, we construct a continuous, infinitely large, and homogeneous surface, where the equivalent circuit-based model of the considered IOS element applies to every single point (unit cell) of the surface. The reflection and transmission coefficients in (\ref{reflection}) and (\ref{transmission}) correspond to this locally equivalent system, in which the incident EM waves are reflected and refracted specularly, i.e., the angles of reflection and refraction coincide with the angle of incidence.
	\end{remark}
\vspace{-0.3cm}
	It is worth mentioning that the equivalent circuit parameters in (\ref{ABCD_matrix}) are determined by projected geometric parameters of the unit cell~\cite{F_2014}. Therefore, the equivalent circuit parameters depend on the angle of incidence, i.e., the inductances, capacitances, resistances, and coupling admittances can be formally written as $(L_1(\theta,\phi),L_2(\theta,\phi),L_3(\theta,\phi))$,  $(C_1(\theta,\phi),C_2(\theta,\phi),C_3(\theta,\phi))$,  $(R_1(\theta,\phi),R_2(\theta,\phi),R_3(\theta,\phi))$,  $(Y_s^{(1)}(\theta,\phi),Y_s^{(2)}(\theta,\phi))$, respectively, where $(\theta,\phi)$ is the angle of incidence. For ease of writing, the angle of incidence $(\theta,\phi)$ is not explicitly stated in (\ref{ABCD_matrix}) and in the subsequent equations. By taking into account the angle-dependence of the equivalent circuit parameters, we can derive the following remark.
	\vspace{-0.3cm}
	\begin{remark}
		\label{reciprocal}
		The reflection and transmission coefficients in (\ref{reflection}) and (\ref{transmission}) depend on the angle of incidence\footnote{For a purely refracting element, the transmission coefficient depends on the angle of incidence of the EM wave.}. Since the proposed reflection-refraction model is based on the locally periodic boundary condition, the angle of reflection of the locally-periodic equivalent of the IOS element is equal to the angle of incidence and, therefore, its dependence is implicitly considered. Thanks to Huygens's principle, however, the whole IOS is capable of realizing anomalous reflections and refractions by adjusting the local reflection coefficient of the IOS elements.  
		
		Besides, the reflection and transmission coefficients depend on the geometric structure of the IOS element, i.e., the geometric period $w_E\times l_E$ of the IOS array, the size $w_M\times l_M$ of the metallic patch, the gap $g$ between the ground and the metallic patch, and the width $w_F$ of the feedline.  
	\end{remark}
\vspace{-0.3cm}
	Compared to existing reflection-refraction models, either for IRSs or IOSs (see, e.g.,~\cite{MFS_arxiv}, Table I), the proposed reflection-refraction model explicitly depends on the hardware implementation of the IOS element, it is electromagnetically consistent under the locally periodic boundary condition, and it allows us to formulate optimization problems in wireless communications that explicitly depend on the tuning circuits for a given geometric structure of the IOS element. The importance of utilizing accurate scattering models for reconfigurable surfaces has recently been analyzed in~\cite{MAAVGMG} with the aid of simulations.
	\vspace{-0.4cm}
	\subsection{Full-wave Validation of the Proposed Circuit-based Reflection-Refraction Model}
	\label{verify}
	\vspace{-0.2cm}

	As mentioned in the previous section, the proposed circuit-based reflection-refraction model of the IOS element depends on a large number of parameters and is derived under the assumption of locally periodic boundary condition. In this section, we aim to validate the proposed model with the aid of full-wave EM simulations. To this end, we consider an implementation of the IOS element, and compare the reflection and transmission coefficients in (\ref{reflection}) and (\ref{transmission}) against those obtained by using a full-wave simulator, i.e., without relying on the circuit-based model in Fig.~\ref{equivalent_circuit} but numerically solving Maxwell's equations.
	
\begin{figure}[!t]
	\centering
	\includegraphics[width=0.4\textwidth]{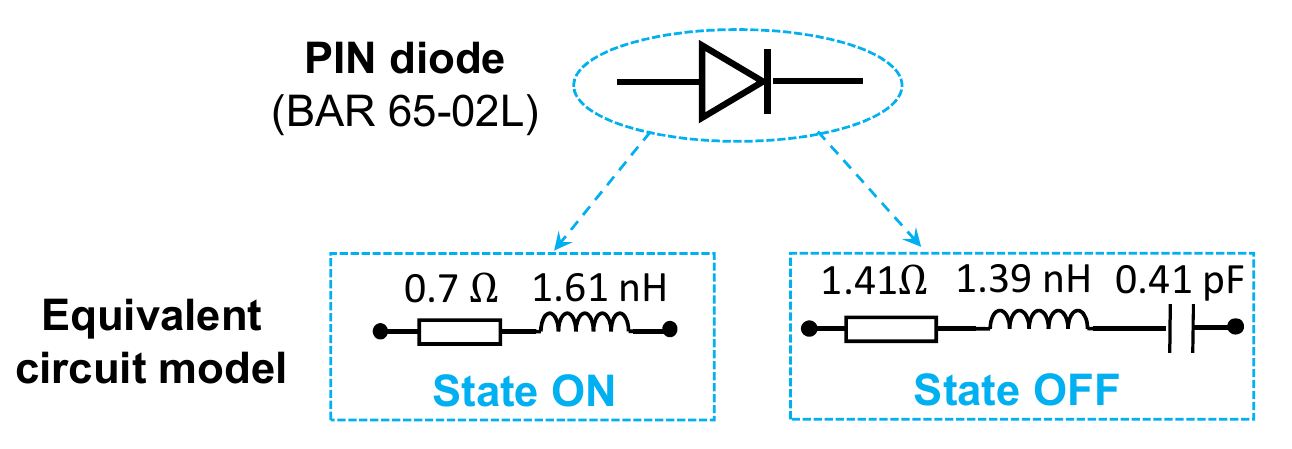}
	\vspace{-3mm}
	\caption{Equivalent circuit model of the PIN diode.}
	\label{equivalent_circuit_PIN_diode}
	\vspace{-7mm}
\end{figure}

	Specifically, we consider an IOS element that operates at $3.6$~GHz and assume that the tunability is ensured by one PIN diode located in the upper metallic plate and by one PIN diode located in the lower metallic plate. The size of the sample IOS element is $2.87 \times 1.42 \times 0.71$ cm$^3$, which implies that the largest side of the IOS element is approximately one-third of the wavelength at $3.6$~GHz. The two layers that constitute the IOS element are separated by a distance of $0.3$~cm. On each layer, a rectangular copper patch whose size is $1.6 \times 1.0$ cm$^2$ (i.e., the size of the patch is approximately $\lambda/5.21 \times \lambda/8.33$ with $\lambda \approx 8.33$~cm at $3.6$~GHz) is printed on a dielectric woven-glass PTFE substrate with permittivity $\epsilon=2.2$ and loss tangent $0.0019@3.6$~GHz. The copper feedline on the other side of the substrate has width equal to $0.04$~cm, and the thickness of the patch and the feedline are $3.9\times 10^{-3}$~cm. {On the boundary of each layer, there is the ground whose width is $0.02$~cm.} The ground is connected to the patch through a BAR 65-02L PIN diode. The equivalent circuit model of the PIN diode is presented in Fig.~\ref{equivalent_circuit_PIN_diode} for completeness~\cite{L_2019}. In the considered example, each IOS element is assumed to be reconfigurable according to two states: both PIN diodes are set to the ON state, i.e., the state (ON, ON), and both diodes are set to the OFF state, i.e., the state (OFF, OFF). Therefore, the set $\mathcal{S}$ has two elements, i.e., $N_S=2$. 

\begin{figure*}[!tpb]
	\centering
	\subfigure[Amplitude vs. frequency under state (ON, ON)]{
		\begin{minipage}[b]{0.45\textwidth}
			\centering
			\includegraphics[width=0.8\textwidth]{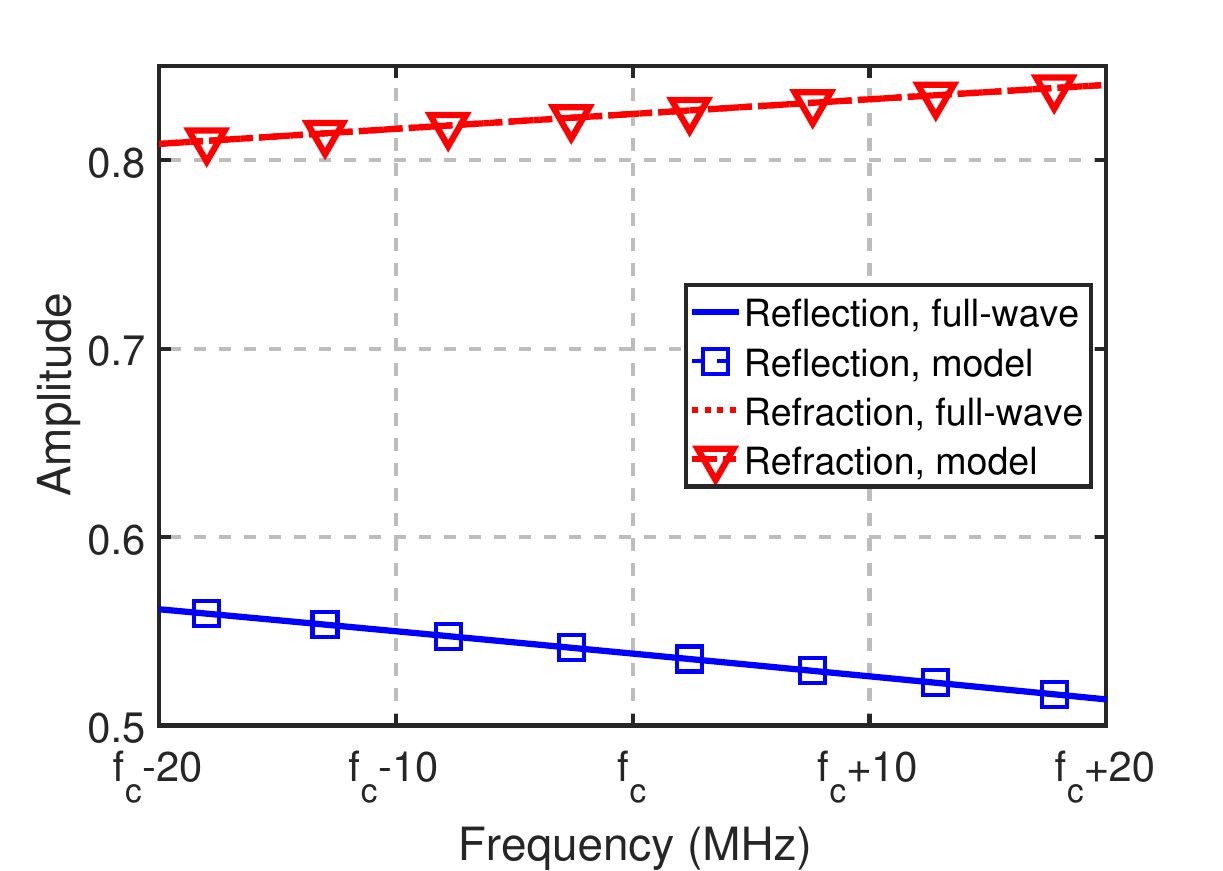}
			\vspace{-0.3cm}
			\label{mag_response_ON}
	\end{minipage}}
	\subfigure[Phase vs. frequency under state (ON, ON)]{
		\begin{minipage}[b]{0.45\textwidth}
			\centering
			\includegraphics[width=0.8\textwidth]{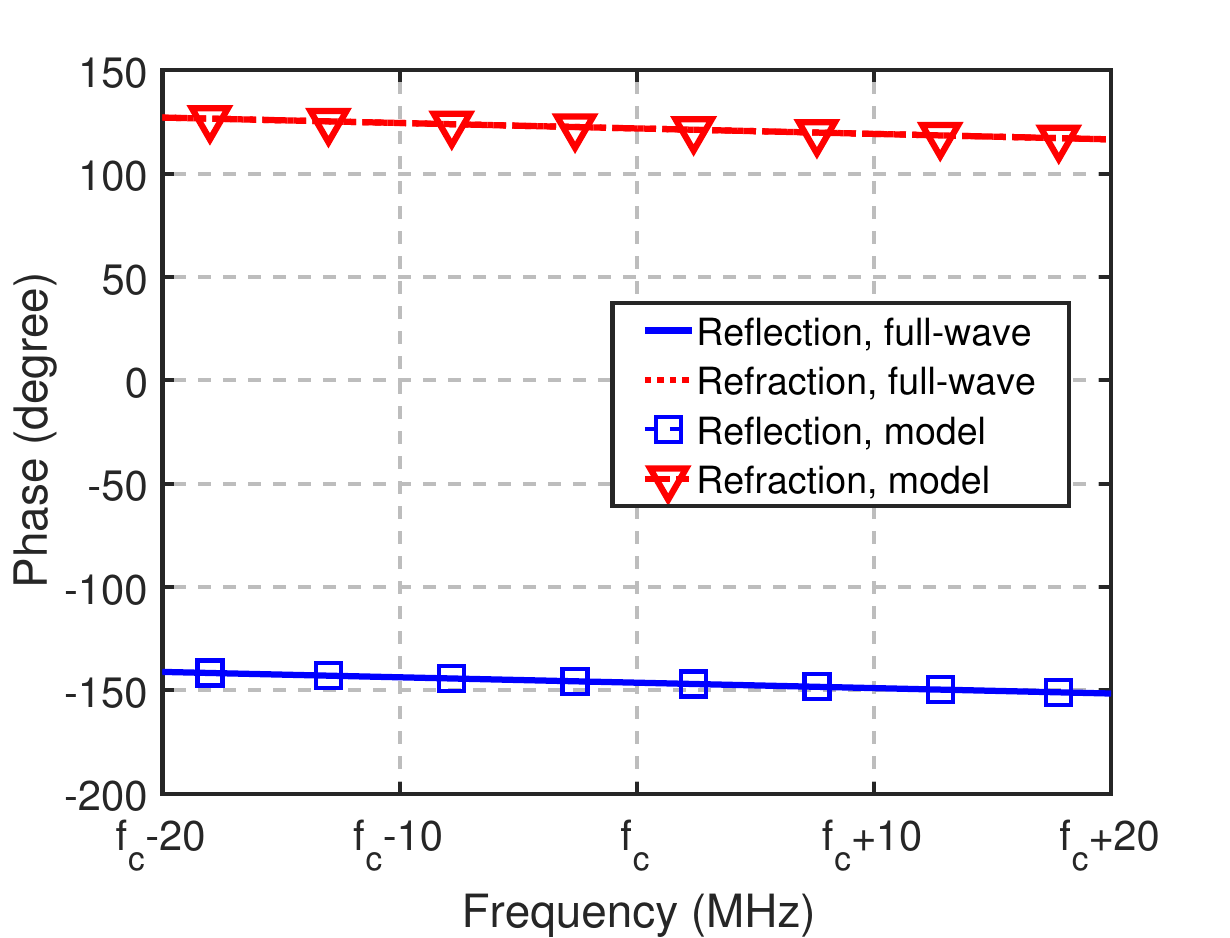}
			\vspace{-0.3cm}
			\label{phase_response_ON}
	\end{minipage}}
	
	\subfigure[Amplitude vs. frequency under state (OFF, OFF)]{
		\begin{minipage}[b]{0.45\textwidth}
			\centering
			\includegraphics[width=0.8\textwidth]{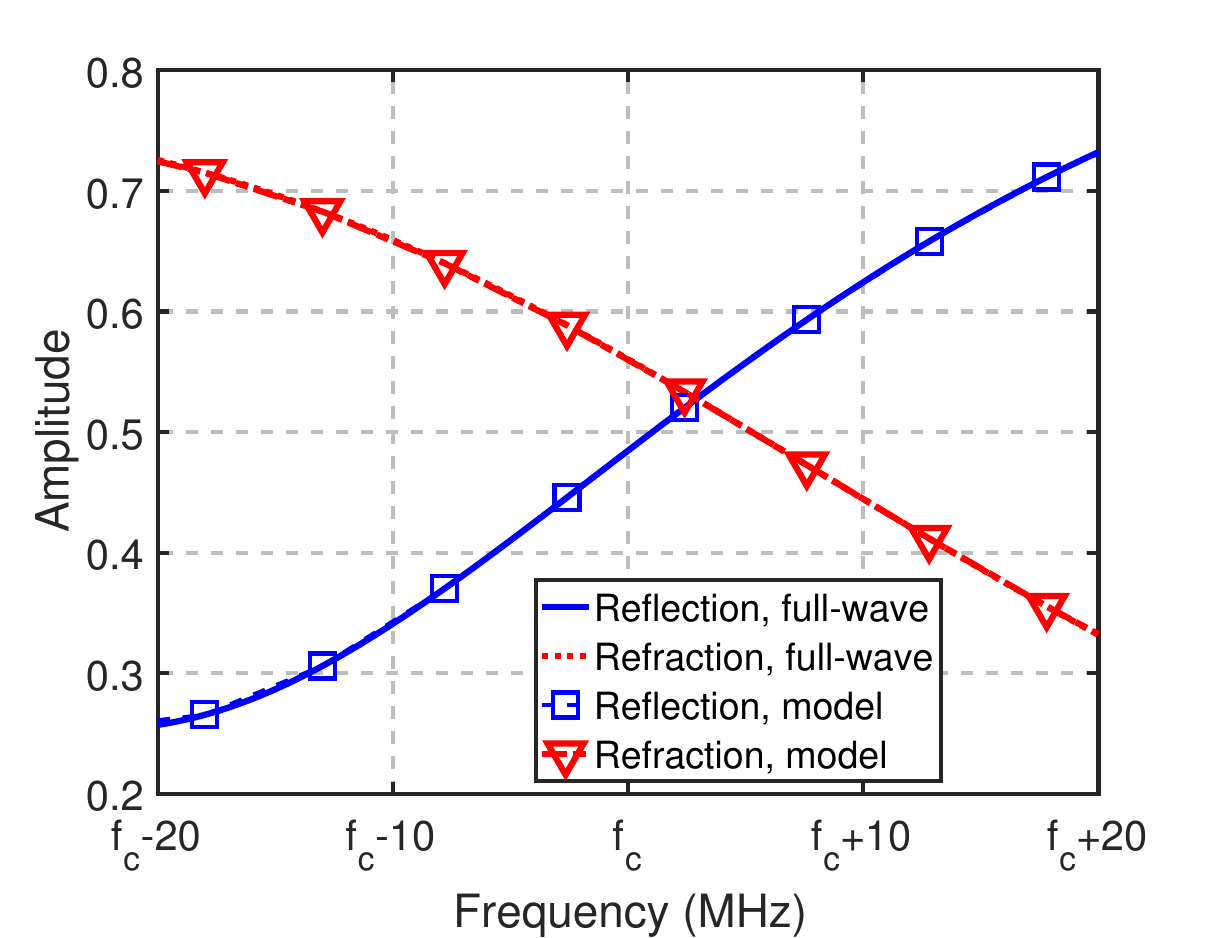}
			\vspace{-0.3cm}
			\label{mag_response_OFF}
	\end{minipage}}
	\subfigure[Phase vs. frequency under state (OFF, OFF)]{
		\begin{minipage}[b]{0.45\textwidth}
			\centering
			\includegraphics[width=0.8\textwidth]{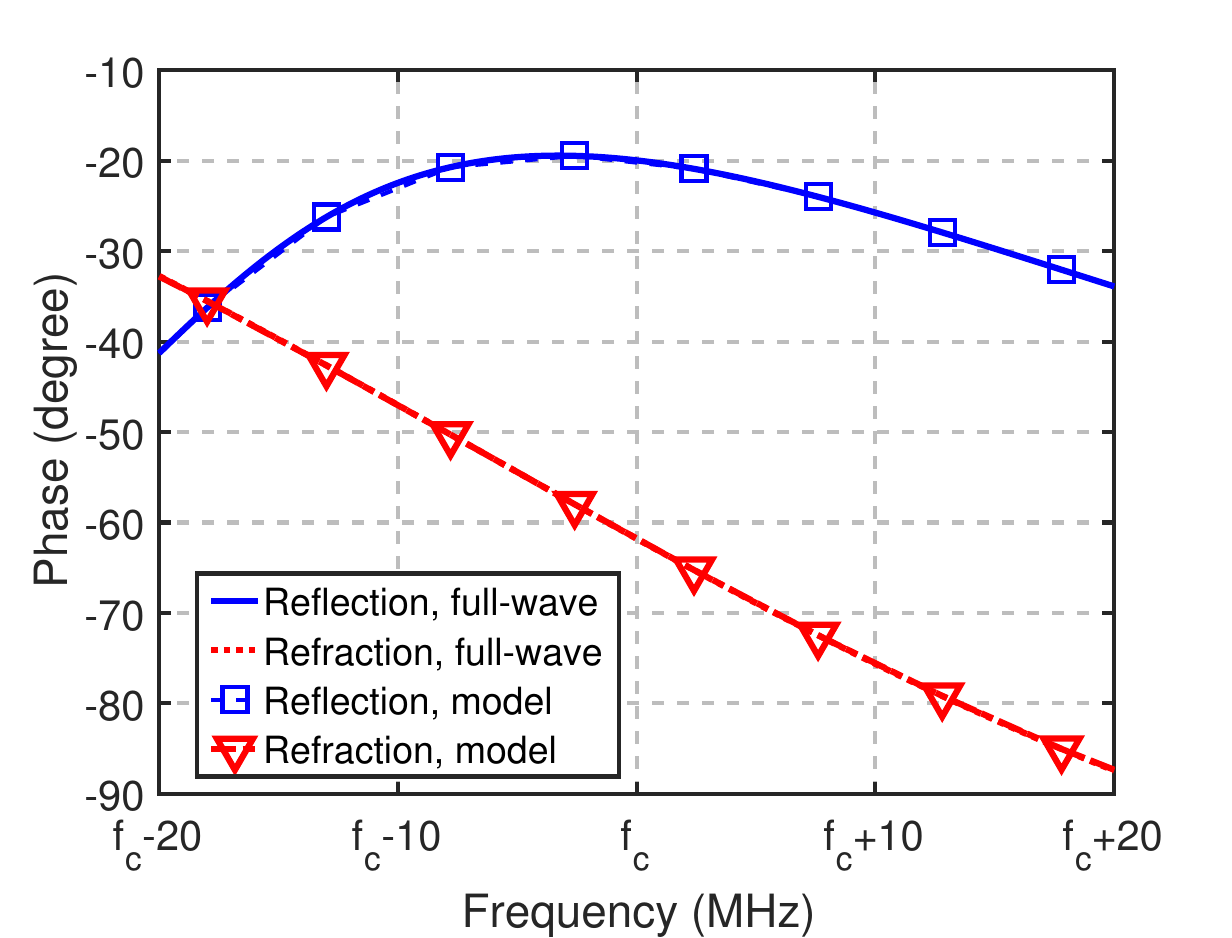}
			\vspace{-0.3cm}
			\label{phase_response_OFF}
	\end{minipage}}
	\vspace{-0.4cm}
	\caption{Amplitude and phase responses of the IOS element versus the operating frequency under normal incidence. Comparison between full-wave simulations and the proposed model. The center frequency is $f_c=3.6$~GHz.}
	\label{response_IOS_ele}
	\vspace{-0.9cm}
\end{figure*}

\begin{figure*}[!tpb]
		\centering
		\subfigure[Amplitude vs. frequency]{
			\begin{minipage}[b]{0.45\textwidth}
				\centering
				\includegraphics[width=0.85\textwidth]{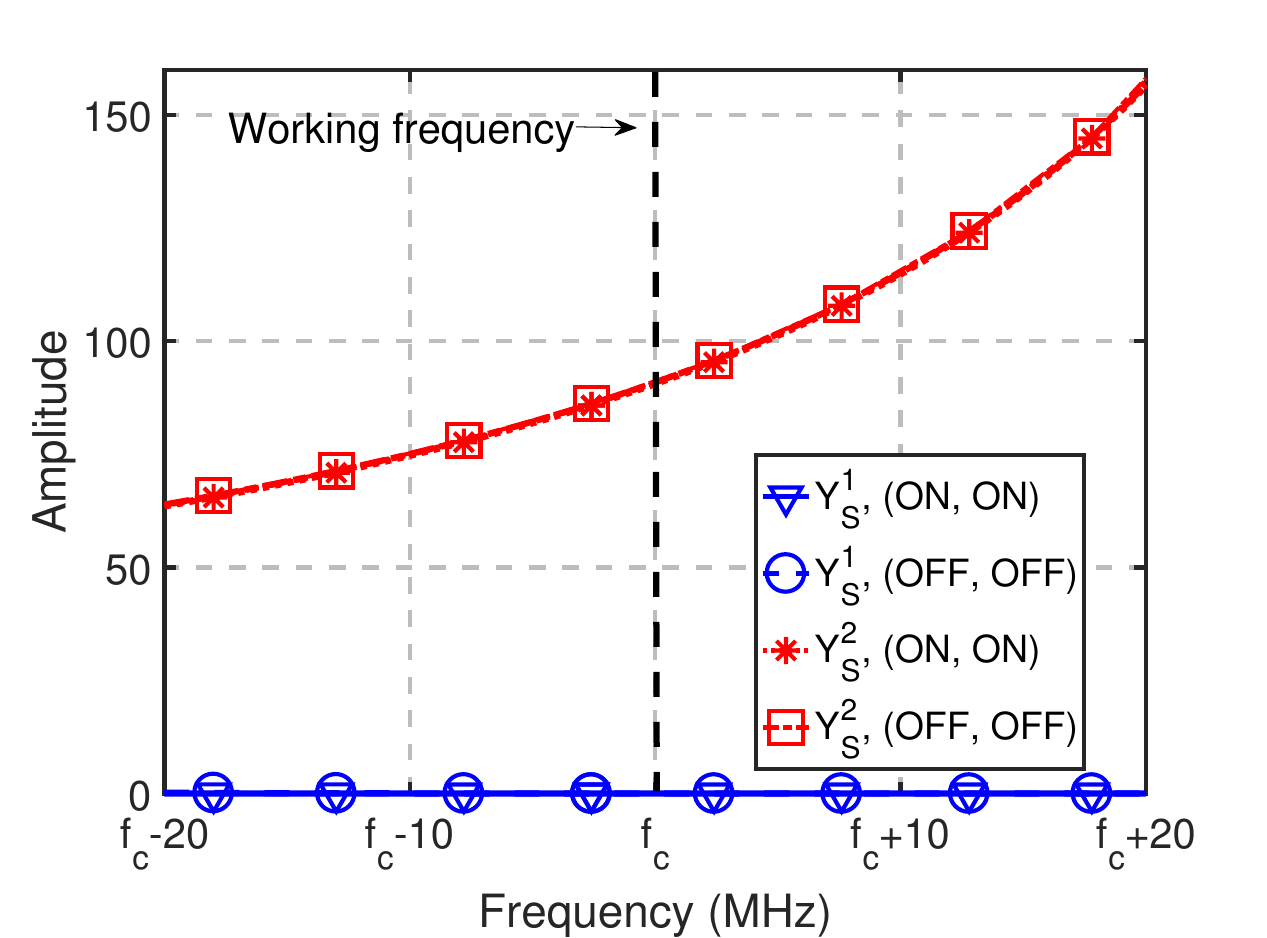}
				\vspace{-0.3cm}
				\label{mag_Y_S}
		\end{minipage}}
		\subfigure[Phase vs. frequency]{
			\begin{minipage}[b]{0.45\textwidth}
				\centering
				\includegraphics[width=0.85\textwidth]{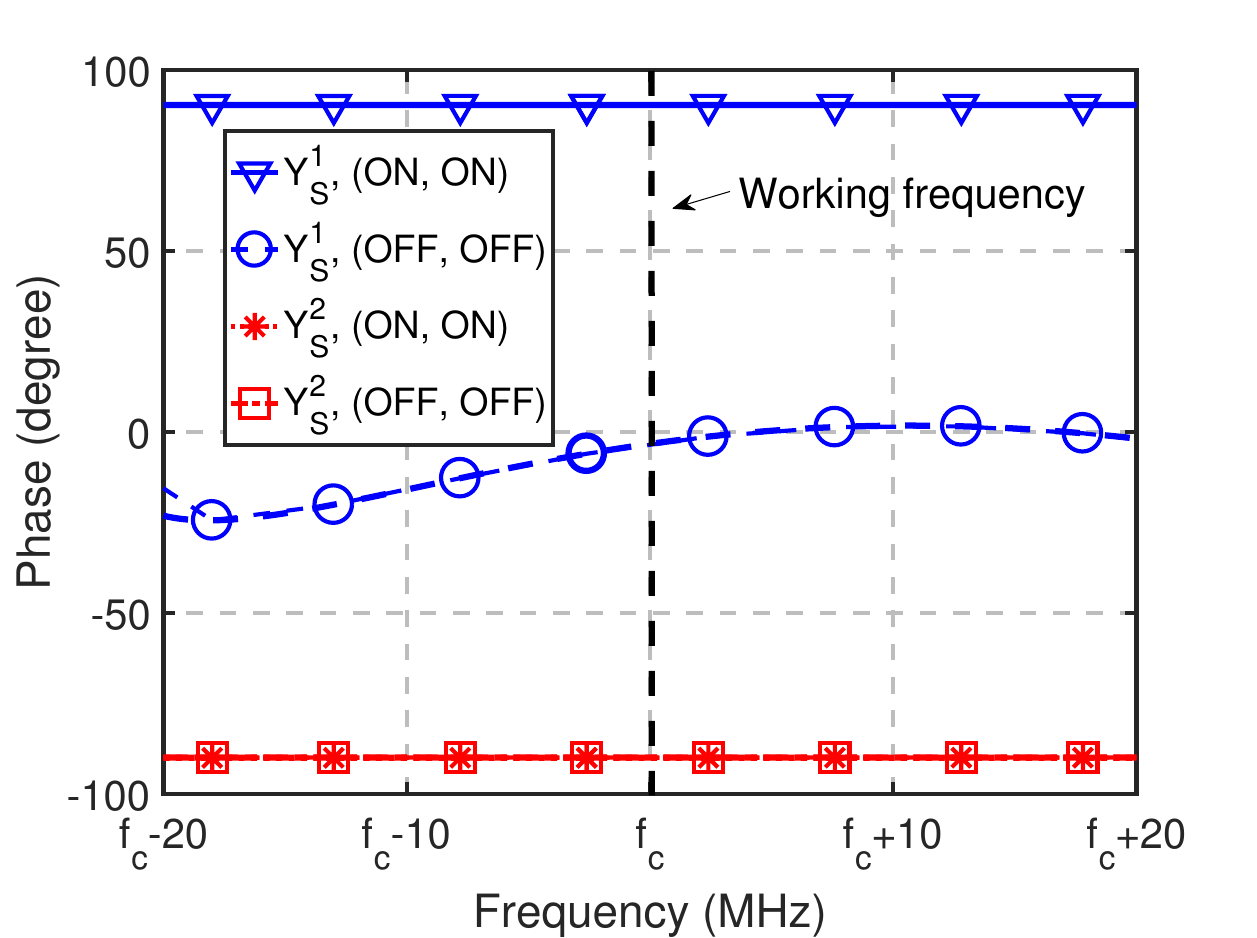}
				\vspace{-0.3cm}
				\label{phase_Y_S}
		\end{minipage}}
		\vspace{-0.5cm}
		\caption{The interlayer coupling admittances $Y_s^{(1)}$ and $Y_s^{(2)}$ of the IOS element under normal incidence. The center frequency is $f_c=3.6$~GHz. $Y_s^{(1)}$ and $Y_s^{(2)}$ depend on the frequency of operation.}
		\label{Y_S}
		\vspace{-0.5cm}
	\end{figure*}

	The full-wave EM simulations are performed by using Microwave Studio and the Transient Simulation Package in the CST software. In the simulations, we assume a plane wave that impinges normally on the IOS element. The EM characterization of the IOS element is performed by imposing the locally periodic boundary condition with Floquet's ports. The distances between the reference planes of the Floquet's ports and the surfaces of the IOS element are $d_1=d_2=0.84\lambda$. In addition, the equivalent circuit model of the PIN diodes is input into the simulation software~\cite{T_2014}. The comparison between the full-wave simulations and the proposed circuit-based model is reported in Fig.~\ref{response_IOS_ele}. \emph{The obtained results show that the proposed model provides a good accuracy for the considered case study.} Specifically, the curves of the circuit-based model are obtained by using (\ref{reflection}) and (\ref{transmission}) with the circuit parameters reported in Table~\ref{equ_para} and Fig.~\ref{Y_S}. These parameters are obtained by matching, through curve fitting, the analytical expressions of the reflection and transmission coefficients in (\ref{reflection}) and (\ref{transmission}) with those obtained from the full-wave EM simulations with CST. The corresponding fitting problem is a quadratic program that is solved by using standard optimization tools. The solution reported in Table~\ref{equ_para} and Fig.~\ref{Y_S} is, of course, not unique. However, this approach substantiates the analytical circuit-based model in (\ref{ABCD_matrix}), (\ref{reflection}), and (\ref{transmission}). As mentioned in previous text, the parameters in Table~\ref{equ_para} and Fig.~\ref{Y_S} are expected to be different under oblique incidence, and the corresponding circuit parameters can be found by using a similar approach. 
	
	\begin{table}[!tpb]
		\setlength{\abovecaptionskip}{0cm}
		\setlength{\belowcaptionskip}{0cm}
		\centering
		\caption{\normalsize{Equivalent circuit parameters of the IOS element under normal incidence}}
		\vspace{0.2cm}
		\label{equ_para}
		\begin{tabular}{|c|c|c|c|c|c|c|c|c|c|}	
			\hline	
			\textbf{Parameter} &  $R_1~(\Omega)$ & $R_2~(\Omega)$ & $R_3~(\Omega)$ & $L_1$~(nH) & $L_2$~(nH) & $L_3$~(nH) & $C_1$~(pF) & $C_2$~(pF) & $C_3$~(pF)\\
			\hline
			\textbf{Value}		& $10^{-3.18}$ & $10^{-3.78}$ & $10^{-7.07}$ & $10^{-3.17}$ & $0.40$ &  $10^{-2.04}$ & $8.03$ & $962.24$ & $209.45$\\
			\hline			
		\end{tabular}
		\vspace{-0.7cm}
	\end{table}

	By direct inspection of Fig.~\ref{response_IOS_ele}, the following observations can be made:
	\begin{itemize}
		\item {Fig.~\ref{mag_response_OFF} indicates that only $55\%$ of the incident EM power is reflected and refracted by the IOS element under state (OFF, OFF), i.e., $|\Gamma_r|^2+|\Gamma_t|^2=0.55$ under state (OFF, OFF).} This is mainly due to the severe losses in the substrate. These losses can be reduced by utilizing a different substrate with a smaller loss tangent and by modifying the geometric structure of the IOS element so that its resonant frequency is significantly different from the working frequency. 
		\item {By comparing Fig.~\ref{phase_response_ON} and Fig.~\ref{phase_response_OFF}, at the center frequency $f_c$, we see that the phase difference between the two states of the IOS element is approximately $180^\circ$ for the refracted signal but only approximately $130^\circ$ for the reflected signal.} The reason is that the difference of the resonant frequency between the two states of the IOS is not large enough.
		\item From Fig.~\ref{response_IOS_ele}, we find that the reflection and refraction responses can be different in terms of amplitude and phase shift, as dictated by (\ref{reflection}) and (\ref{transmission}). According to  (\ref{reflection}) and (\ref{transmission}), specifically, this implies that the power ratio and the phase shift difference between the reflected and refracted signals are determined by the geometric structure and hardware implementation of the IOS element. The current implementation of the IOS element does not allow us to adjust the power ratio and the phase shift difference between the reflected and refracted signals. The design of a reconfigurable IOS element with these  tunable characteristics is postponed to a future implementation.
	\end{itemize}


\begin{figure*}[!t]
	\centering
	\subfigure[Amplitude vs. the angle of incidence]{
		\begin{minipage}[b]{0.45\textwidth}
			\centering
			\includegraphics[width=0.85\textwidth]{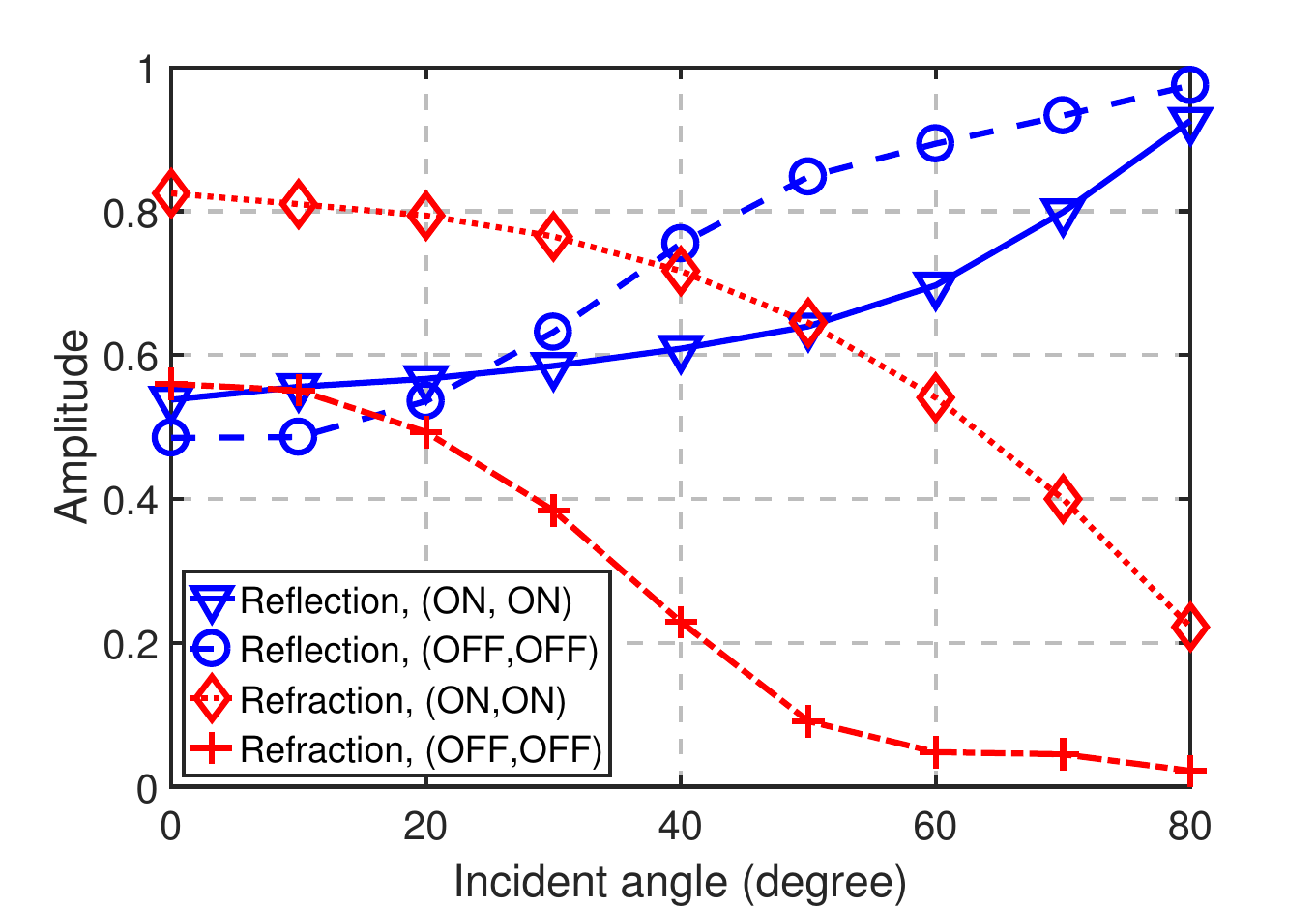}
			\vspace{-0.2cm}
			\label{incident_angle_amp}
	\end{minipage}}
	\subfigure[Phase vs. the angle of incidence]{
		\begin{minipage}[b]{0.45\textwidth}
			\centering
			\includegraphics[width=0.85\textwidth]{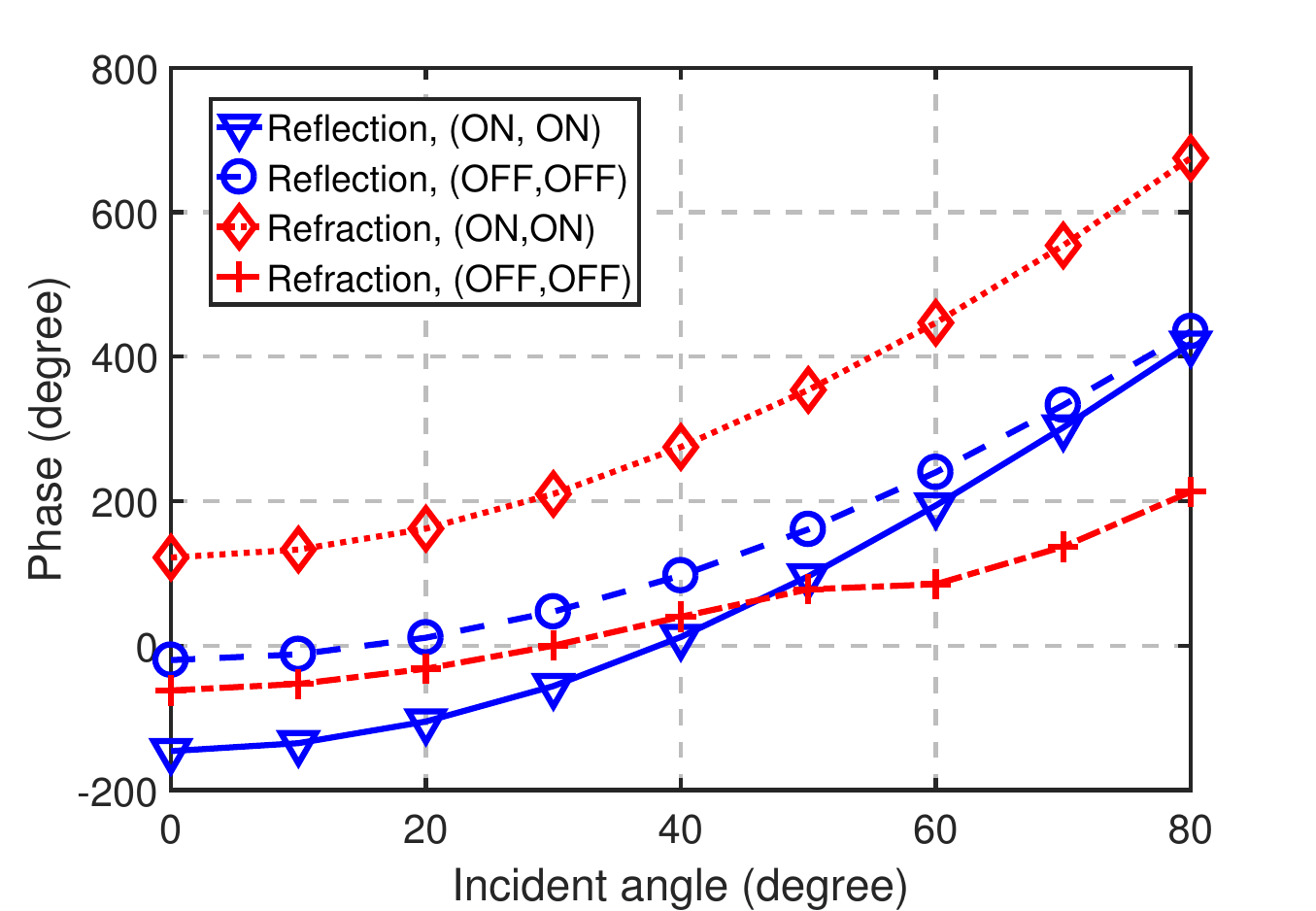}
			\vspace{-0.2cm}
			\label{incident_angle_phase}
	\end{minipage}}
	\vspace{-4mm}
	\caption{Full-wave simulations of the reflection and transmission coefficients vs. the angle of incidence. The operating frequency is $3.6$~GHz.}
	\label{incident_angle}
	\vspace{-9mm}
\end{figure*}

	As mentioned in Remark~\ref{reciprocal}, the reflection and transmission coefficients in (\ref{reflection}) and (\ref{transmission}) depend on the angle of incidence of the EM waves. This remark is validated in Fig.~\ref{incident_angle} with the aid of full-wave EM simulations. Figure~\ref{incident_angle} confirms that the reflection and transmission coefficients depend on the angle of incidence. Also, the angle-dependence is different for the (ON, ON) and (OFF, OFF) states. Specifically, the following observations can be made from Fig.~\ref{incident_angle}:
	\begin{itemize}
		\item From Fig.~\ref{incident_angle_amp}, we see that the amplitude of the reflection and transmission coefficients increases and decreases, respectively, with the angle of incidence. This implies that, in general, the amount of power that is reflected and refracted depends on the angle of incidence.
		\item From Fig.~\ref{incident_angle_phase}, we see that the designed IOS element offers, however, better performance in refraction than in reflection. The reason is that the phase difference between the two states of the IOS element is approximately equal to $180^\circ$ for the refracted wave, while it tends to shrink for the reflected wave as the angle of incidence increases. This implies that, for the reflected waves, it is not possible to impose the required phase shift on the waves that impinge on the IOS from a wide angle of incidence. Good reflection capabilities are obtained for angles of incidence that are less than $40^\circ$.
	\end{itemize}

\vspace{-3mm}
	\section{Full-dimensional Beamforming}
	\vspace{-2mm}
	\label{bf}
	Based on the proposed reflection-refraction model, we introduce in Section~\ref{framework} a general framework for hybrid beamforming, which is specifically tailored for full-dimensional communications. In Section~\ref{metric}, we then introduce different performance metrics for optimizing and evaluating full-dimensional communications.
	\vspace{-0.6cm}
	\subsection{General Framework for Hybrid Beamforming}
	\vspace{-0.1cm}
	\label{framework}
	We first introduce the basic concepts for hybrid beamforming. In an IOS-assisted network, the BS and the IOS perform \emph{digital beamforming} and \emph{analog beamforming}, respectively. Specifically, the BS first encodes multiple data streams via a digital beamformer $\bm{V}_D$. Then, the encoded signal is up-converted to the desired carrier frequency and is transmitted through $K_t$ transmit antennas. When the signals from the BS impinge upon the IOS, each IOS element reflects and refracts the incident signal by applying a certain phase shift to the signal. This operation is referred to as analog beamforming, and it is usually realized through a wireless control channel that the BS utilizes to configure the amplitude and phase responses of the IOS elements. Reconfigurable surfaces that do not need any control channels have recently been proposed as well, but they are out of the scope of this paper~\cite{AFVMX}. In full-dimensional communications, therefore, the digital beamformer at the BS and the analog beamformer at the IOS can be jointly optimized for achieving the optimal performance.
	
	At the first sight, this hybrid beamforming design is similar to IRS-assisted communications. However, IOS-aided systems face several new design challenges. First, the reflection and transmission coefficients of the IOS can be asymmetric, which implies that the channel models of the UEs on two sides of the IOS are in general different. Therefore, the hybrid beamforming schemes utilized for IRSs cannot be directly applied to IOSs. Second, differently from an IRS, an IOS is deployed to support full-dimensional communications on both sides of the surface. Therefore, the design of the hybrid beamforming needs to account for both the reflection and transmission coefficients, and the coverage or rate requirements on both sides of the surface.
	
	To cope with these design challenges, we formulate a joint BS digital beamforming and IOS analog beamforming optimization problem. Let $G(\bm{V}_D,\bm{s})$ denote the generic performance metric of interest (to be detailed in the next section), where $\bm{V}_D$ and $\bm{s}$ denote the BS digital beamforming matrix and the IOS analog beamforming vector, respectively\footnote{The general function $G(\bm{V}_D,\bm{s})$ can be any performance metric mentioned in Section~\ref{metric}.}. Specifically, the state of the $m$-th IOS element is denoted by $s_m$. Also, the reflection and transmission coefficients of the $m$-th IOS element when the signal transmitted by the $k$-th Tx antenna impinges upon the element are represented by $\Gamma_r^{m,k}$ and $\Gamma_t^{m,k}$, respectively, and the angle of incidence is denoted by $(\theta_{in}^{m,k},\phi_{in}^{m,k})$. Then, the optimization problem can be formulated as
{\setlength\abovedisplayskip{0cm}
	\setlength\belowdisplayskip{0cm}
	\begin{subequations}\label{h_bf}
	\begin{align}
	&\max_{\bm{V}_D,\bm{s}}~G(\bm{V}_D,\bm{s}),\label{obj}\\
	s.t.~&{\gamma_j\ge \gamma_0},\label{QoS}\\
	& [\bm{V}_D\bm{V}_D^{\mathrm{H}}]_{k,k} \le P_T,\label{cons_power}\\
& (\Gamma_r^{m,k},\Gamma_t^{m,k})\in\left\{\left(\Gamma_r(\theta_{in}^{m,k},\phi_{in}^{m,k},s_m),\Gamma_t(\theta_{in}^{m,k},\phi_{in}^{m,k},s_m)\right)\right\}_{s_m\in \mathcal{S}}, \forall m,\label{cons_state}
	\end{align}
\end{subequations}
	where {the constraint in (\ref{QoS}) ensures the desired quality of service for each UE by making sure that the received signal-to-interference-and-noise ratio $\gamma_j$ is larger than a predetermined threshold $\gamma_0$}, the constraint in (\ref{cons_power}) ensures that the transmit power of each antenna at the BS cannot exceed the maximum value $P_T$, and the constraint in (\ref{cons_state}) indicates that the reflection and transmission coefficients of the $m$-th IOS element can only take discrete values that belong to the finite set $\mathcal{S}$.} It is worthwhile noting that the constraint in (\ref{cons_state}) depends on the angle of incidence. Specifically, the reflection and transmission coefficients of each IOS element belong to a different feasible set because the angle of incidence is not usually the same for all the IOS elements. Only when the transmitter is in the far-field of the IOS, the curvature of the incident wave across the surface of the IOS can be ignored, and the angle of incidence can be assumed to be the same across the entire IOS. Efficient algorithms to solve the problem in (\ref{h_bf}) can be found in~\cite{HSBYMDLZH} and~\cite{SHBYMZL_2020}. {Specifically, for computational efficiency, the digital beamformer and the IOS configuration are optimized alternatively. For the digital beamforming subproblem, the channels are fixed because the IOS configuration is given. Therefore, the problem is a well-known digital beamforming problem, and thus the zero-forcing digital beamformer is adopted to obtain a near optimal solution~\cite{FDBETF_2013}. For the analog beamforming subproblem, the states of the IOS elements are optimized one-by-one to further reduce the computational complexity, i.e., one at a time, the state of only one IOS element is optimized to maximize the system performance in (\ref{obj}) given the states of the other IOS elements.}

	\vspace{-0.4cm}
	\subsection{Performance Metrics}
	\label{metric}
	In this section, we summarize relevant performance metrics that can be considered for optimizing full-dimensional communications. These performance metrics correspond to the objective function in (\ref{obj}). The considered performance metrics are divided into two categories: \textit{beam pattern related} and \textit{data rate related}.
	\subsubsection{Beam pattern related metrics}
	We focus our attention on the beam pattern corresponding to the $j$-th UE, which refers to the normalized power that is scattered by the IOS as a function of a generic direction of observation under the assumption that the IOS is optimized for maximizing the power towards the location of the $j$-th UE. Four metrics related to the optimization of the beam pattern are considered.
	\begin{enumerate}[itemindent=0em, label=$\bullet$]
		\item \textbf{Beam direction}: The direction of a beam is defined as the direction of the main lobe.
		\item \textbf{Half-power beamwidth}~(HPBW): The HPBW identifies the range of directions towards which the radiated power is no smaller than half of the power radiated towards the beam direction~\cite{WQWLQ_2018}. A narrower beamwidth indicates a larger beamforming gain, which can lead to a stronger desired signal. Also, a narrow beamwidth reduces the inter-user interference among closely-spaced UEs.
		\item \textbf{Sidelobe level}~(SLL): The SLL refers to the power difference of the largest sidelobe with respect to the main lobe~\cite{Q_2018}. The SLL is an effective metric to quantify the amount of interference generated towards undesired directions. The smaller the SLL, the smaller the interference.
		\item \textbf{Scattering efficiency}: The scattering efficiency is defined as the percentage of power in the main beam with respect to the total scattered power~\cite{VBSF_2021}. The larger the scattering efficiency, the stronger the desired signal and the smaller the interference.
	\end{enumerate}
	Therefore, a well designed IOS is characterized by a narrow HPBW, a low SLL, and a high scattering efficiency, while steering the beams towards the desired UEs. In an IOS, this needs to be jointly ensured for the reflected and refracted beams. Specifically, the beam pattern of an IOS is directly determined by the reflection and transmission coefficients in (\ref{reflection}) and (\ref{transmission}), respectively, and the digital beamforming matrix $\bm{V}_D$. In mathematical terms, the beam pattern corresponding to the $j$-th UE can be formulated as
	{\setlength\abovedisplayskip{0cm}
		\setlength\belowdisplayskip{0cm}
	\begin{align}
	\label{pattern}
	F_j(\theta,\phi)=\Big|\sum_k E_k(\theta,\phi)V_{D_{k,j}}\Big|^2,
	\end{align}
	where $(\theta, \phi)$ is the far-field angle under which the IOS views the $j$-th UE, $V_{D_{k,j}}$ is the $k$-th element in the $j$-th row of the digital beamforming matrix $\bm{V}_D$ at the BS, and $E_k(\theta,\phi)$ is the electric field scattered by the IOS towards the direction $(\theta,\phi)$ under the assumption that only the $k$-th antenna of the BS is activated.} Specifically, $E_k(\theta,\phi)$ is determined by the configuration of the IOS, i.e., by the reflection and transmission coefficients in (\ref{reflection}) and (\ref{transmission}), and it is independent of the BS digital beamformer. In the following, $E_k(\theta,\phi)$ is referred to as the \emph{reference beam pattern}. The proposed model in (\ref{pattern}) is validated through experimental results in Appendix~\ref{app_model_beam}. 
	
	\subsubsection{Data rate}
	The data rate is a typical performance metric of interest in wireless communications. In a multi-user scenario, it is defined as the sum-rate over all the UEs. Specifically, the data rate of $j$-th UE is defined as~\cite{SHBYMZHL}
	{\setlength\abovedisplayskip{0cm}
		\setlength\belowdisplayskip{0cm}
	\begin{align}
	\label{data_rate}
	R_j=\log_2\left(1+\Big|C_j\sum\limits_{k}V_{D_{k,j}}h_{k,j}\Big|^2\Big/\Big(\sigma^2+\sum\limits_{j'\neq j}\Big|C_{j}\sum\limits_{k}V_{D_{k,j'}}h_{k,j'}\Big|^2\Big)\right).
	\end{align} 
	where $C_j$ accounts for the joint impact of the RF chains at the BS and the $j$-th UE, $\sigma^2$ is the variance of the zero-mean additive white Gaussian noise~(AWGN) at the $j$-th UE, and $h_{k,j}$ is the channel gain from the $k$-th antenna of the BS to the $j$-th UE, which includes the impact of the IOS. As far as the IOS is concerned, we consider the same model as in~\cite{SHBYZHL_2021}. More precisely, we assume that the IOS elements are partitioned into $M$ groups and that the elements belonging to the same group are set to the same state\footnote{In general, the grouping strategy needs to jointly consider the system performance and the implementation complexity, which can be broadly summarized as follows: 1) Grouping adjacent IOS elements rather than dispersed elements is beneficial to simplify the design of the feedline network, especially if each group contains the same number of elements; 2) Minimizing the number of groups while fulfilling the required system performance, so as to find a desirable tradeoff between system performance and implementation complexity.}. This is a typical approach for realizing reconfigurable surfaces at a reduced complexity. Then, the IOS-aided channel $h_{k,j}$ can be formulated as~\cite{J_2020}
	\begin{align}
	\label{uncoupled}
	h_{k,j}=h_{k,j}(s_1=0,\dots,s_{M}=0)+\sum_{m=1}^{M}s_m\Delta h_{k,j}^{(m)},
	\end{align} 	
	where $s_m$ denotes the state of the $m$-th group of IOS elements,} $\Delta h_{k,j}^{(m)}$ denotes the channel difference when $s_m$ changes from the state 0, i.e., the (OFF, OFF) state, to the state 1, i.e., the (ON, ON) state, and the other groups of IOS elements are kept to the state 0. In mathematical terms, we have
	{\setlength\abovedisplayskip{0cm}
		\setlength\belowdisplayskip{0cm}
	\begin{align}
	\label{channel_change}
	\Delta h_{k,j}^{(m)}=h_{k,j}(s_1=0,\dots,s_m=1,\dots,s_{M}=0)-h_{k,j}(s_1=0,\dots,s_{M}=0).
	\end{align}

	\section{Experimental Prototype}
	\label{prototype}
	In order to experimentally validate the proposed reflection-refraction model for the IOS, and to evaluate the performance of full-dimensional beamforming, we realized an IOS hardware prototype that is described in this section.}

	
	\vspace{-0.6cm}
	\subsection{Implementation of the IOS}
	\vspace{-.2cm}


	As shown in Fig.~\ref{hardware_module}, the implemented IOS consists of 640 reconfigurable elements. The structure and the electromagnetic characterization of each IOS element was introduced in Section~\ref{verify}. To configure the states of the IOS elements in order to manipulate the incident waves, we utilize a field-programmable gate array (FPGA) platform that mimics the function of the IOS controller. A software program is pre-loaded in the FPGA, and the states of the IOS elements are changed automatically by controlling the FPGA. The FPGA is a Cyclone IV EP4CE10F17C8 platform that is manufactured by Intel Altera corporation. The development board that contains the FPGA and various interfaces is manufactured by ALINX. The hardware description language used for programming the FPGA is Verilog.
	
\vspace{-5mm}
	\subsection{Hardware Modules of the Prototype}
	\vspace{-2mm}
	\begin{figure*}[!t]
		\centering
		\includegraphics[width=0.60\textwidth]{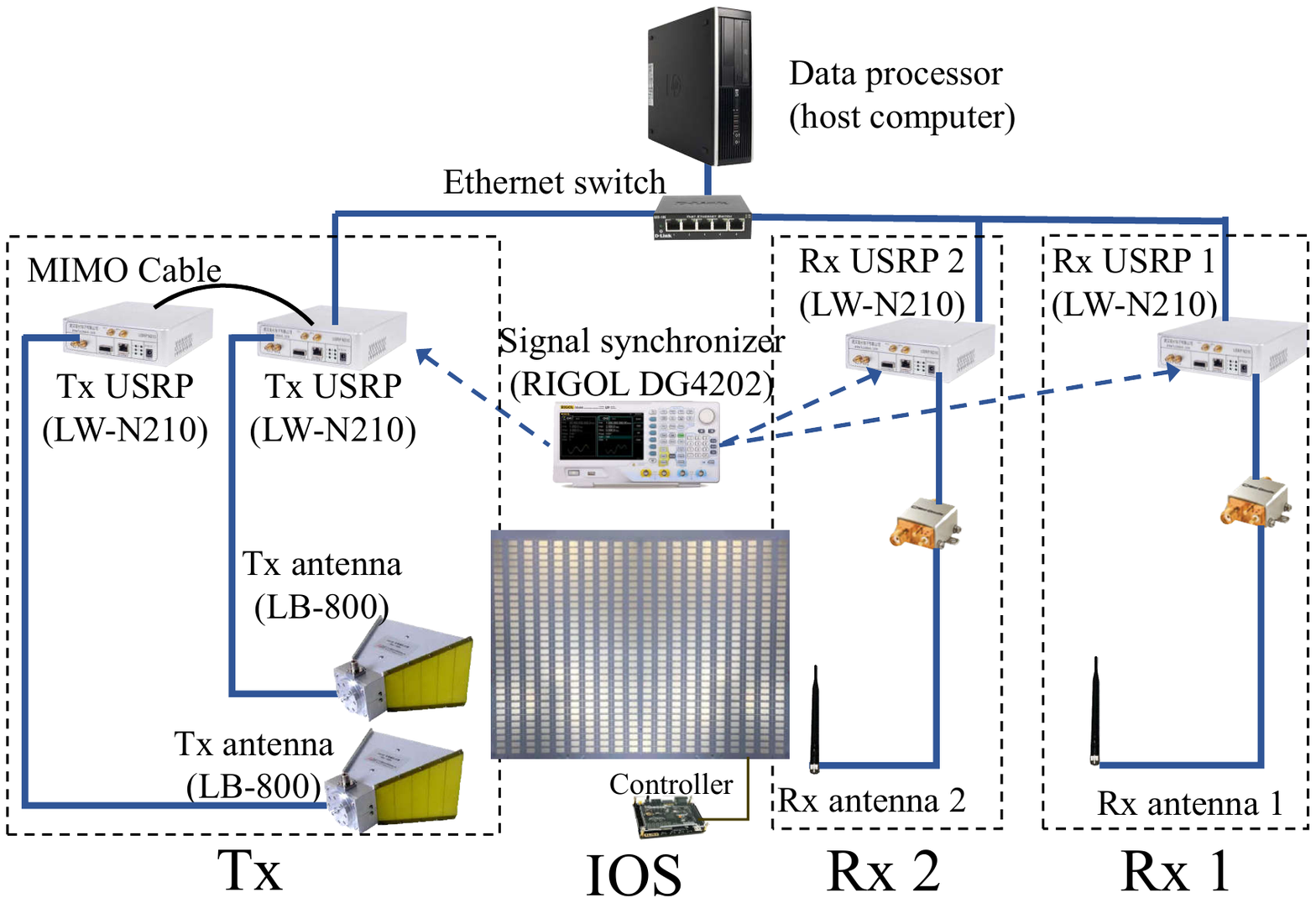}
		\vspace{-5mm}
		\caption{Hardware modules of the implemented IOS prototype.}
		\label{hardware_module}
		\vspace{-9mm}
	\end{figure*}
	As shown in Fig.~\ref{hardware_module}, the developed IOS hardware platform comprises the following modules.
	\begin{itemize}
	\item \emph{Transmitter:} The transmitter~(Tx) includes two transmit elements. Each transmit element is constituted by a universal software radio peripheral~(USRP) platform. Specifically, an LW N210 USRP platform with an SBX-LW120 RF daughterboard is used. The output port of each USRP is connected to a directional double-ridged horn antenna whose gain is $12.5$~dBi at the working frequency of $3.6$~GHz. The antenna is manufactured by A-INFO Corporation and the part number is LB-880. A MIMO cable is utilized to  synchronize the transmission of data between the two USRP platforms. Each USRP platform implements the carrier modulation\footnote{The implemented IOS-based prototype does not perform digital modulation and coding, since we evaluate the theoretical rate in (\ref{data_rate}).}, power amplification, and filtering by using a GNU radio software development kit~\cite{OHA-2015}. The maximum transmit power is $20$~dBm.
	
	\item \emph{Receiver:} Two receivers~(Rxs) are implemented. Each receive element is constituted by one USRP, one low-noise amplifier~(LNA), and one cylindrical antenna. The signal received by the antenna is first amplified by the LNA whose part number is ZX60-43-S+, and it is then sent to the USRP for the carrier demodulation and the baseband processing. The gain of the antenna and the gain of the LNA are $3$~dBi and $15.07$~dB, respectively, at $3.6$~GHz. 
	
	\item \emph{Signal synchronizer:} {To accurately detect the relative phases of the received signals with respect to the transmitted signals, a clock source that outputs the same clock signal to the Tx USRP and Rx USRPs is utilized, and a RIGOL DG4202 signal source generator is used to generate the same pulse-per-second~(pps) signal to the USRPs. The synchronization of the clock and pps signals can reduce the time-variance of the carrier phase difference between the Tx and Rx USRPs, which improves the accuracy for measuring the channel phase\footnote{Fig.~\ref{hardware_module} illustrates an indoor application where Rx 1 and Rx 2 on two sides of the IOS are synchronized with the Tx.}.}
	
	\item \emph{Ethernet switch:}  The Ethernet switch connects the transmitter and the receivers to a host computer to exchange information. Specifically, through the Ethernet switch, the host computer sends the control signals to the transmitter and receiver units, and the receivers return their received signals to the host computer. The bandwidth	of the switch is about 1 GHz.
	
	\item \emph{Data processor:} The data processor is a host computer that controls the transmitter and the receivers through a software program written in Python. {The digital beamforming is operated by the data processor as well}.
\end{itemize}
As a whole, the different modules of the IOS prototype (in particular the USRP and the FPGA) are coordinated through a software program written in Python that runs on the host computer. Specifically, the host computer configures the USRP platforms through the GNU Radio software development kit. The FPGA, in turn, transforms the input from the host computer into a control signal for the IOS with the aid of a software program written in Verilog.

\section{Measurement Procedures}
\vspace{-.1cm}
\label{metrics_process}
In this section, we introduce the measurement procedures that are utilized to characterize the beam pattern (Section~\ref{meas_pattern}) and the data rate (Section~\ref{meas_data_rate}).

\vspace{-5mm}
\subsection{Beam Pattern}
\vspace{-1mm}
\label{meas_pattern}
In order to experimentally evaluate the performance of the IOS in terms of reradiation capabilities, the beam pattern of a generic UE in (\ref{pattern}) needs to be first measured and then analyzed in terms of different performance metrics. To this end, the reference beam patterns that characterize the IOS as a function of its configuration need to be measured. In this section, we describe, without loss of generality, the procedure for measuring the reference beam pattern that corresponds to the receiver Rx 1 in Fig.~\ref{hardware_module}.

\begin{figure*}[!t]
	\centering
	\includegraphics[width=0.5\textwidth]{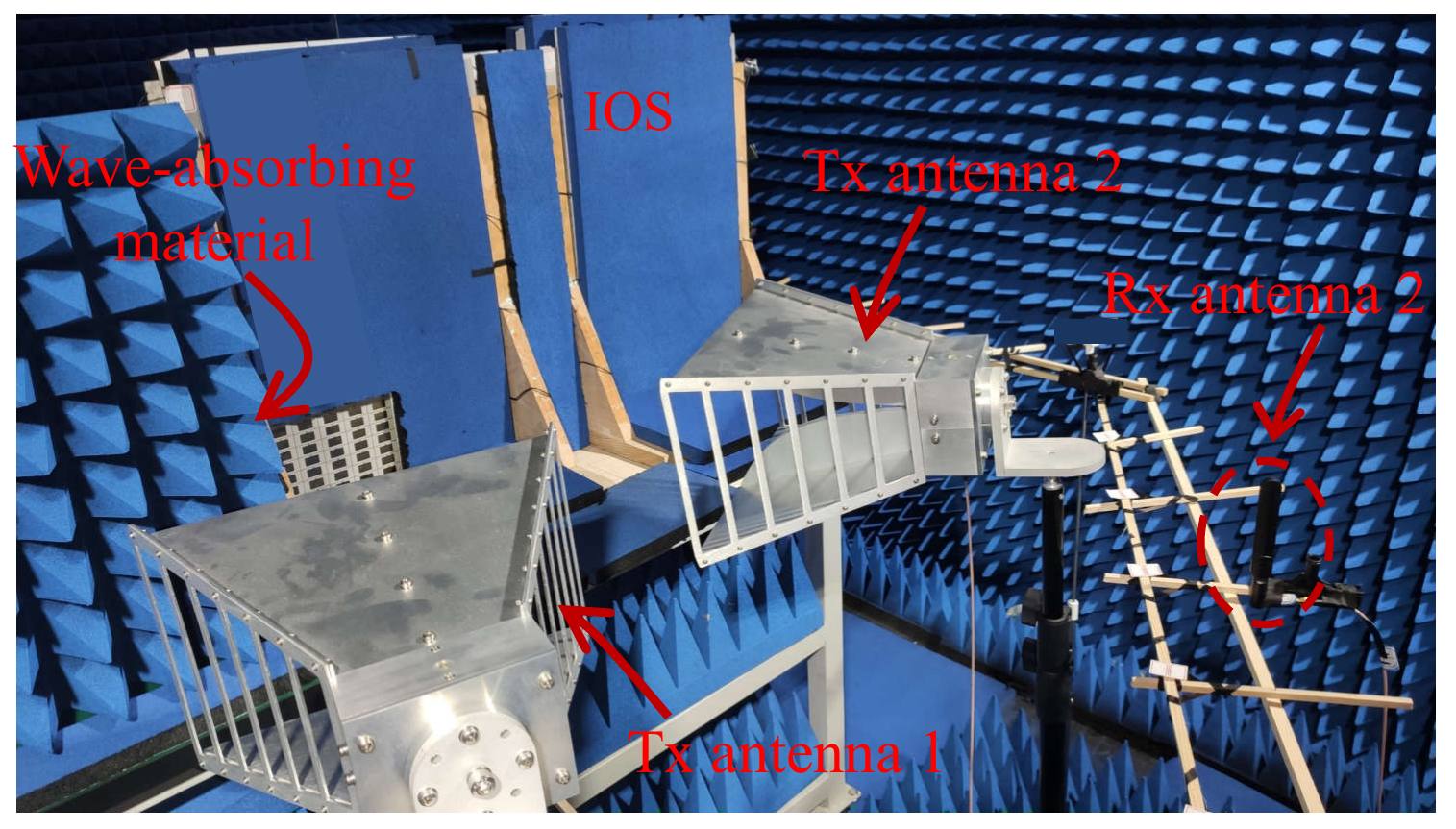}
	\vspace{-2mm}
	\caption{Environment for measuring the beam patterns.}
	\label{exp_setting}
			\vspace{-11mm}
\end{figure*}

To measure the reference beam patterns for different IOS configurations, we deploy the transmitter, the receiver, and the IOS as illustrated in Fig.~\ref{exp_setting}. Since we consider the receiver Rx 1, the transmitter is configured so that the symbols $x_{2}$ intended for Rx $2$ is set equal to zero, i.e., $x_{2}=0$, in order to avoid interference. For simplicity of measurement, the transmitted signal $x_1$ intended for Rx $1$ is set equal to $x_1=1$, i.e., a sequence of all ones is transmitted to Rx $1$. By definition, as mentioned, the reference beam pattern $E_k(\theta,\phi)$ is obtained by turning the $k$-th transmit antenna on and by turning all the other transmit antennas off.

To conduct the measurements, we carefully setup the measurement environment as illustrated in Fig.~\ref{exp_setting}. The IOS prototype is deployed within a room whose walls are made of aluminum in order to block environmental EM interference. To avoid environmental scatterings, the wall of the room is covered with wave absorbing material. To measure only the beam scattered by the IOS, the direct link between the transmitter and the receiver needs to be strongly attenuated. To this end, the transmitter is placed closer to the IOS as compared to the distance between the Rx and the IOS, and some wave-absorbing material is placed in the middle of the room. To measure the reradiation pattern of the IOS as a function of the angle of observation, we build a circular measurement platform that is located parallel to the ground\footnote{For simplicity of measurement, we focus our attention on the beam within the horizontal plane instead of the three-dimensional beam pattern. Therefore, a measurement platform parallel to ground is sufficient.}. Then, the reference field pattern $E_k(\theta,\phi)$ is measured by deploying Rx $1$ in different locations on the measurement platform. Once the reference beam patterns are measured, the beam pattern of the IOS is obtained from (\ref{pattern}), where the BS digital beamforming matrix in (8) is determined by utilizing the algorithm described in Section~\ref{framework}.

\vspace{-5mm}
\subsection{Data Rate}
\label{meas_rate}
\label{meas_data_rate}
\vspace{-3mm}
In order to measure the data rate, we utilize the following approach. First, we measure the channel parameters, and then we insert them in (\ref{data_rate})-(\ref{channel_change}) for obtaining the data rate. This procedure is convenient because the acquisition of the channel parameters is needed for optimizing the IOS configuration and the BS digital beamforming matrix.


Based on (\ref{data_rate})-(\ref{channel_change}), the estimation of $C_jh_{k,j}$ is sufficient to compute the rate\footnote{The implemented IOS-based prototype does not perform signal detection, since we evaluate the theoretical rate in (\ref{data_rate}).}. Specifically, $C_jh_{k,j}$ are extracted by setting the IOS to a specific configuration and by measuring the received signal. Let $k$ and $k'$ denote the indices of the two transmit antennas in Fig.~\ref{hardware_module}, {and let $j'$ denote the index of the Rx other than Rx $j$.} Then, the signal received at the $j$-th receiver (i.e., Rx $j$) can be expressed as~\cite{YBHMLL_2022}
{\setlength\abovedisplayskip{0cm}
	\setlength\belowdisplayskip{0cm}
\begin{align}
\label{sig_rec}
y_j=C_jV_{D_{k,j}}h_{k,j}x_j+C_jV_{D_{k',j}}h_{k',j}x_j+C_{j}\sum\limits_{k_t\in\{k,k'\}}V_{D_{k_t,j'}}h_{k_t,j'}x_{j'}+n_j
\end{align}
{where $x_j$ and $x_{j'}$ denote the intended signal for Rx $j$ and Rx $j'$, respectively}, and $n_j$ is the receive AWGN.} The first two terms on the right-hand side of (\ref{sig_rec}) represent the desired signals from the $k$-th and the $k'$-th antennas of the transmitter, respectively. On the other hand, the third term denotes the inter-user interference. Based on (\ref{sig_rec}), the channel parameter $C_jh_{k,j}$ are acquired by extracting the first term from the received signal $y_j$, and by dividing it by $x_jV_{D_{k,j}}$. To this end, the inter-user interference $C_{j}\sum_{k_t\in\{k,k'\}}V_{D_{k_t,j'}}h_{k_t,j'}x_{j'}$ needs to be mitigated by setting $\{V_{D_{k,j'}},V_{D_{k',j'}}\}=0$, and the signal $C_jV_{D_{k',j}}h_{k',j}$ needs to be mitigated by setting $V_{D_{k',j}}=0$. The impact of the AWGN $n_j$ is weakened by increasing the gain $C_j$ of the RF chains, so as to enhance the signal-to-noise-ratio. 

To estimate the rate in (\ref{data_rate}), the variance $\sigma^2$ of the received AWGN is necessary as well. To measure the variance $\sigma^2$, the transmitter is turned off and the unbiased estimator $\sum_{u=1}^{U}(y_u-\bar{y})^2/(U-1)$ is utilized~\cite{RM_2012}, where $y_u$ is one sample of (\ref{sig_rec}), and $\bar{y}$ is the average values from $U$ collected samples. Once the channel parameters and the noise variance are obtained, the data rate in (\ref{data_rate}) is computed by setting the configuration of the BS according to the hybrid beamforming algorithm formulated in Section~\ref{framework}.

\vspace{-5mm}
\section{Experimental Results}
\vspace{-2mm}
\label{verify_all}
In this section, we describe experimental results obtained with the implemented IOS prototype, as well as we evaluate the performance of full-dimensional beamforming and verify the theoretical findings.


	\begin{figure*}[!t]
		\centering
		\includegraphics[width=0.3\textwidth]{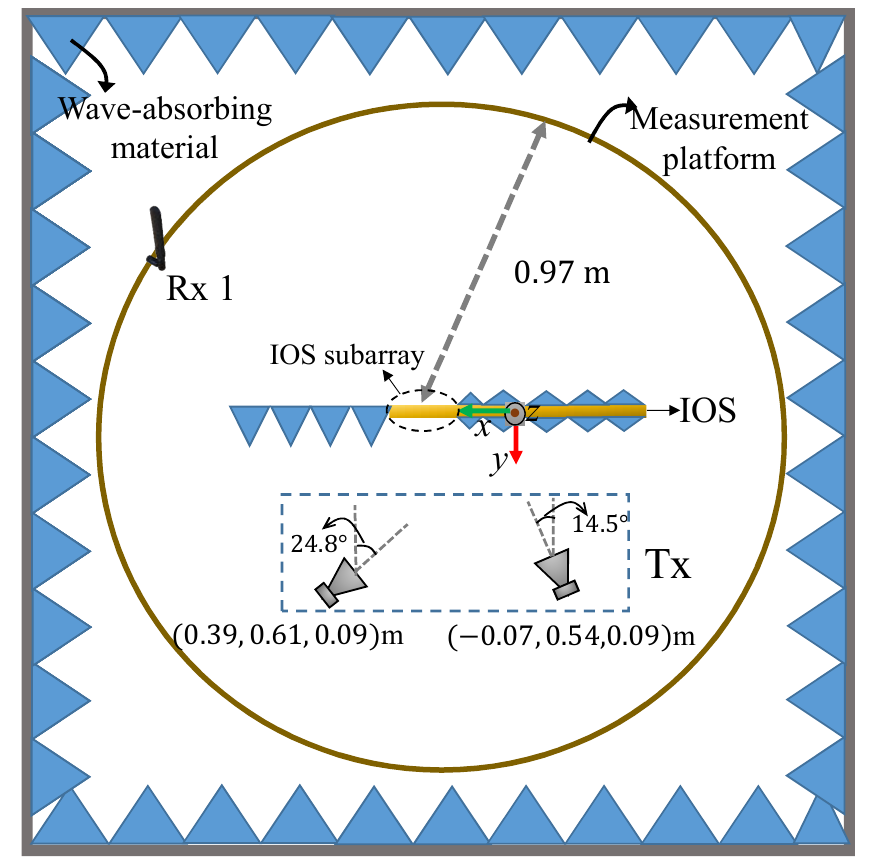}
			\vspace{-2mm}
		\caption{Experimental layout for measuring the beam patterns.}
		\label{pattern_layout}
				\vspace{-7mm}
	\end{figure*}

\vspace{-5mm}
	\subsection{Analysis of the Beam Pattern}
	\label{exp_radiation_pattern}
\begin{figure*}[!tpb]
		\centering
		\subfigure[]{
			\begin{minipage}[b]{0.3\textwidth}
				\centering
				\includegraphics[width=0.9\textwidth]{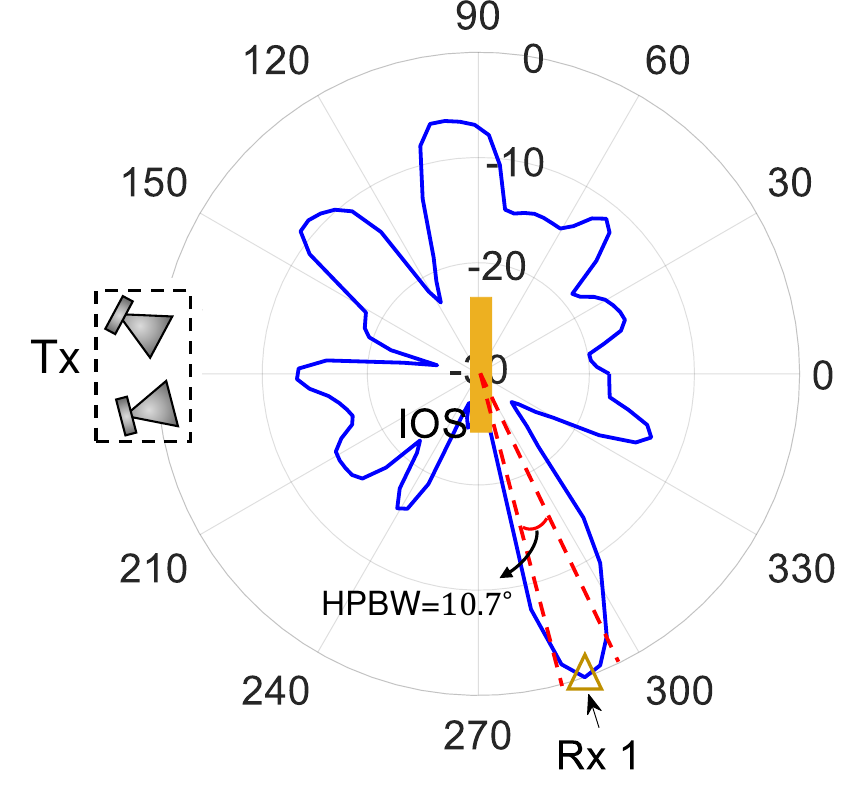}
						\vspace{-0.2cm}
				\label{dir_289}
		\end{minipage}}
		\subfigure[]{
			\begin{minipage}[b]{0.3\textwidth}
				\centering
				\includegraphics[width=0.9\textwidth]{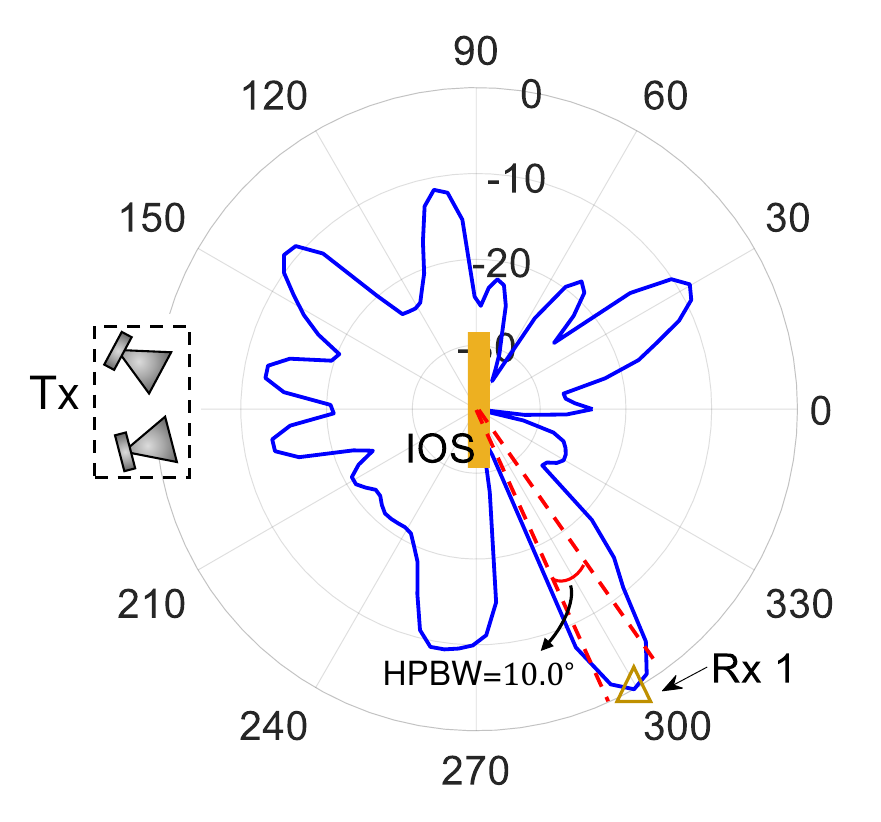}
				\vspace{-0.2cm}
				\label{dir_299}
		\end{minipage}}	 	
		\subfigure[]{
			\begin{minipage}[b]{0.32\textwidth}
				\centering
				\includegraphics[width=0.95\textwidth]{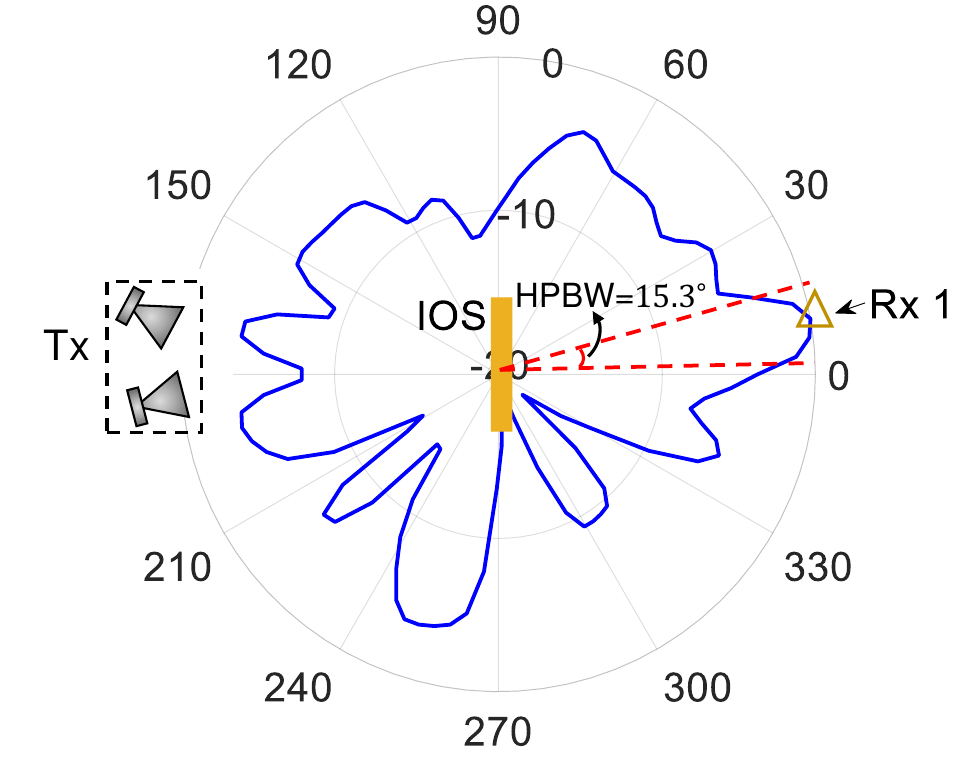}
				\vspace{-0.2cm}
				\label{dir_10}
		\end{minipage}}
	\vspace{-0.4cm}
		\caption{Scattered beams when the IOS is configured to point towards the refraction region. In (a), (b), and (c), the beam is directed towards $289^\circ$, $299^\circ$, and $10^\circ$, respectively.}
		\label{beam_transmission}
		\vspace{-0.3cm}
	\end{figure*}

\begin{figure*}[!tpb]
	 	\centering
		\subfigure[]{
	 		\begin{minipage}[b]{0.3\textwidth}
	 			\centering
	 			\includegraphics[width=0.9\textwidth]{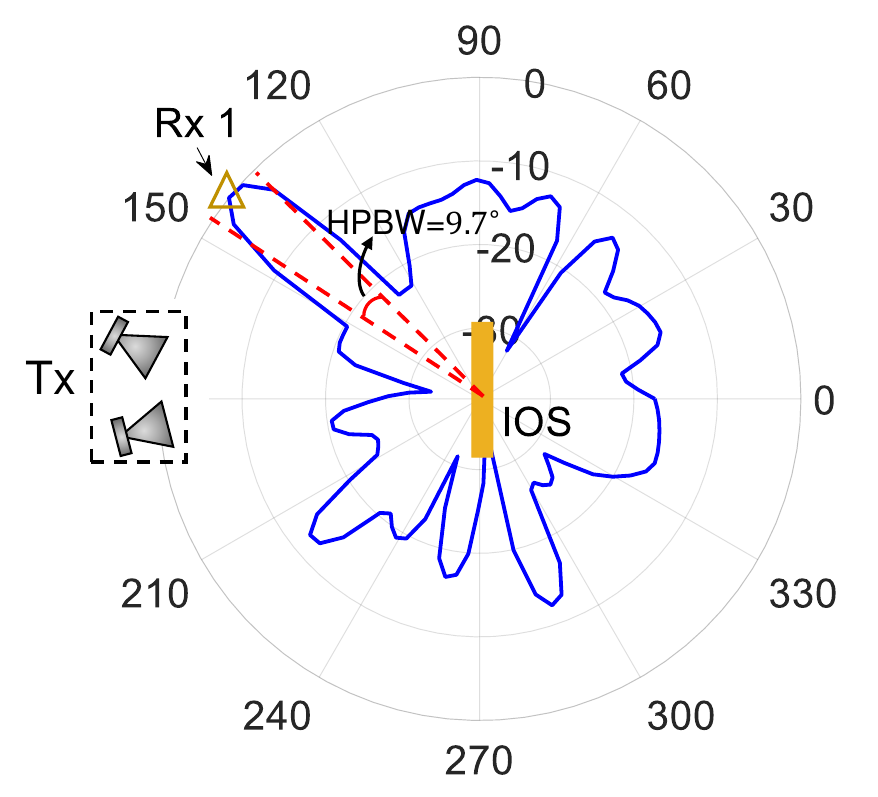}
	 			\vspace{-0.2cm}
	 			\label{dir_141}
	 	\end{minipage}}
		\subfigure[]{
	 		\begin{minipage}[b]{0.3\textwidth}
	 			\centering
	 			\includegraphics[width=0.9\textwidth]{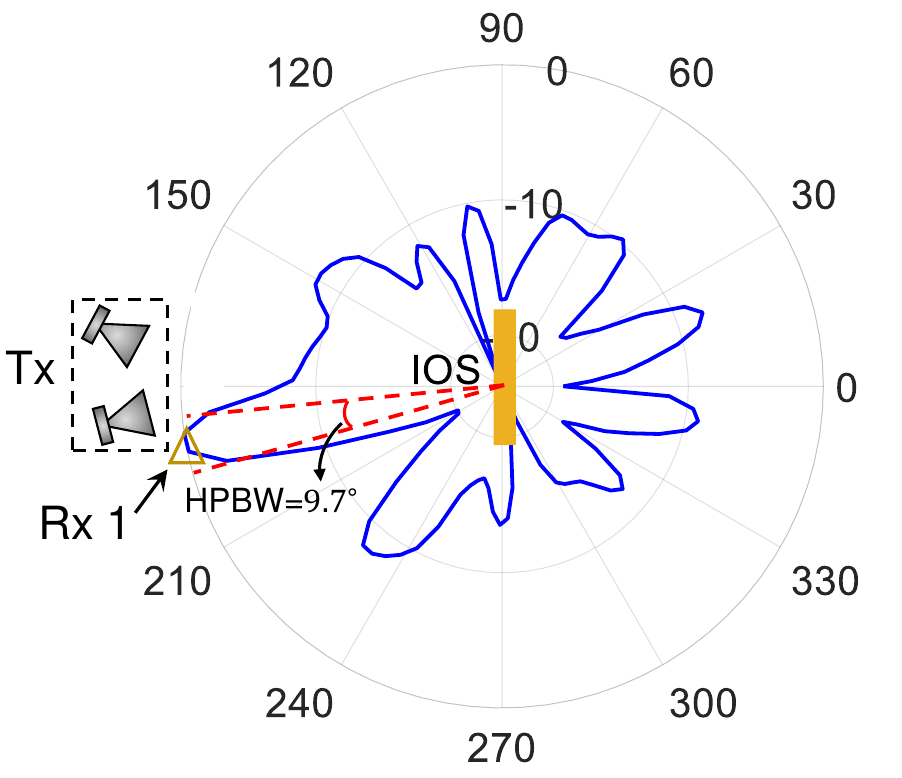}
	 			\vspace{-0.2cm}
	 			\label{dir_188}
	 	\end{minipage}}	 	
		\subfigure[]{
	 		\begin{minipage}[b]{0.3\textwidth}
	 			\centering
	 			\includegraphics[width=0.9\textwidth]{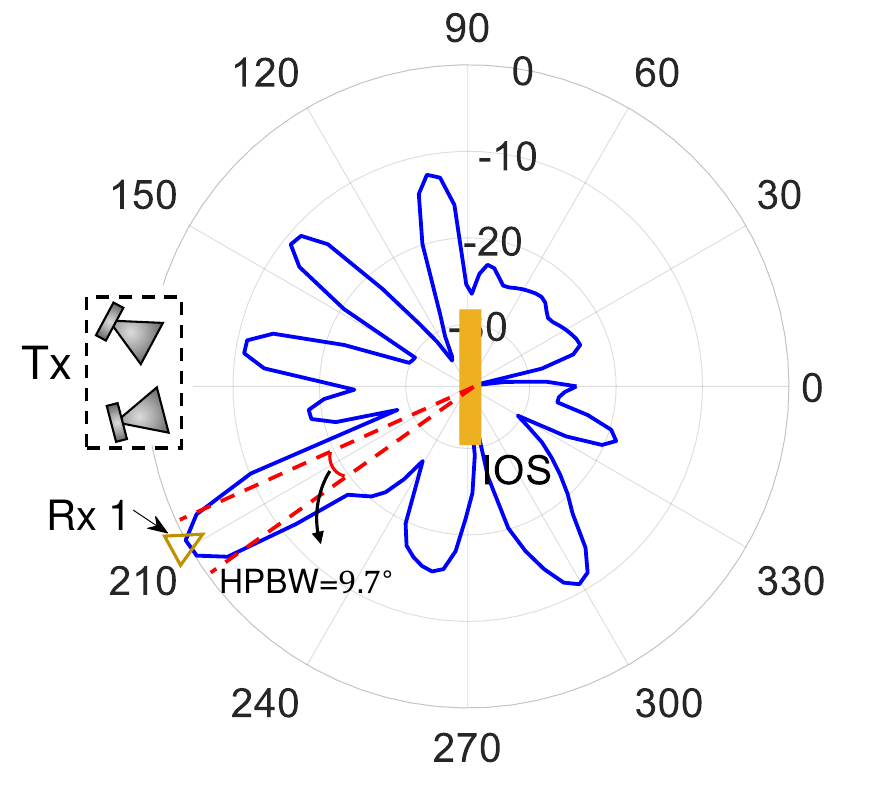}
	 			\vspace{-0.2cm}
	 			\label{dir_208}
	 	\end{minipage}}
 	\vspace{-0.4cm}
	 	\caption{Scattered beams when the IOS is configured to point towards the reflection region. In (a), (b), and (c), the beam is directed towards $141^\circ$, $188^\circ$, and $208^\circ$, respectively.}
	 	\label{beam_reflection}
	 	\vspace{-3mm}
	 \end{figure*}

\begin{figure*}[!t]
		\centering
		\subfigure[Rx 1 is in the reflection region.]{
			\begin{minipage}[b]{0.45\textwidth}
				\centering
				\includegraphics[width=0.98\textwidth]{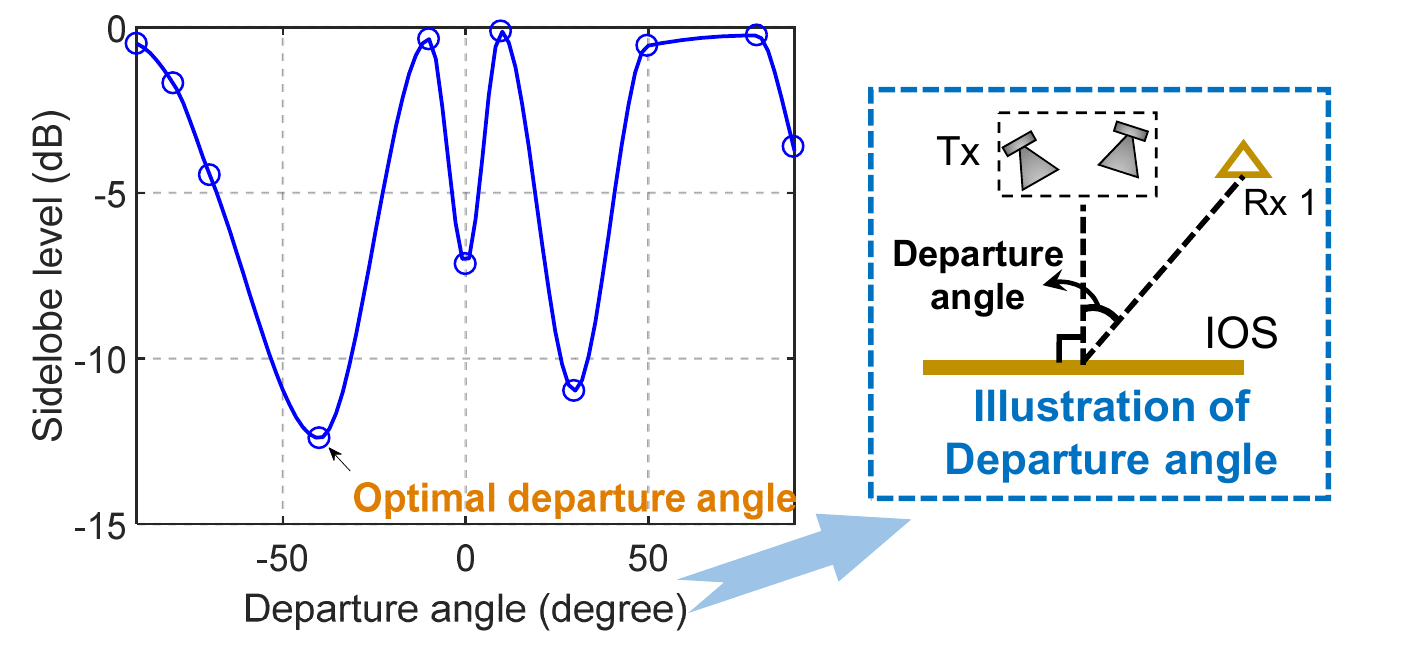}
				\vspace{-0.3cm}
				\label{SLL_reflection_region}
		\end{minipage}}
		\subfigure[Rx 1 is in the refraction region.]{
			\begin{minipage}[b]{0.45\textwidth}
				\centering
				\includegraphics[width=0.98\textwidth]{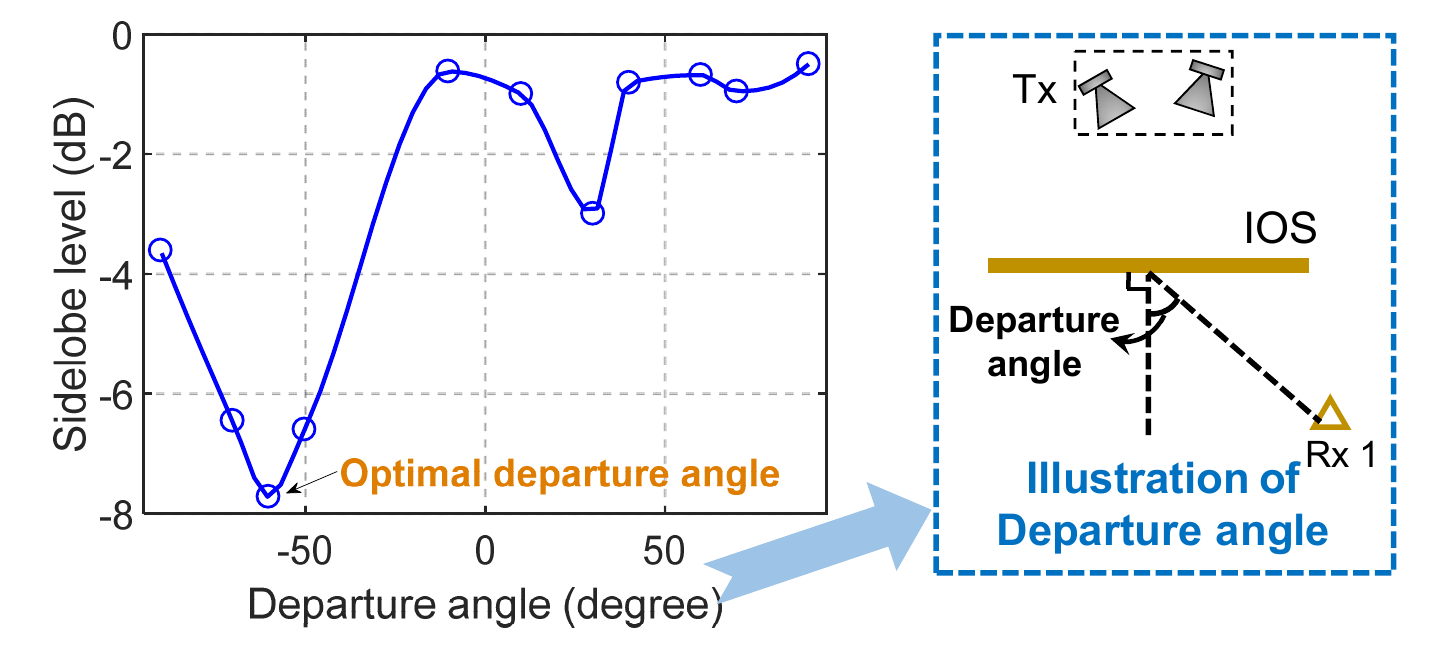}
				\vspace{-0.3cm}
				\label{SLL_refraction_region}
		\end{minipage}}
		\vspace{-6mm}
		\caption{Sidelobe level vs. the angle of departure from the IOS, when the scattered beam points towards the desired receiver (Rx 1). The angle of departure coincides with the angle of reflection or the angle of refraction, depending on the location of Rx 1.}
		\label{SLL}
		\vspace{-3mm}
	\end{figure*}

\begin{figure*}[!t]
	\centering
	\subfigure[Rx 1 is in the reflection region]{
		\begin{minipage}[b]{0.45\textwidth}
			\centering
			\includegraphics[width=0.7\textwidth]{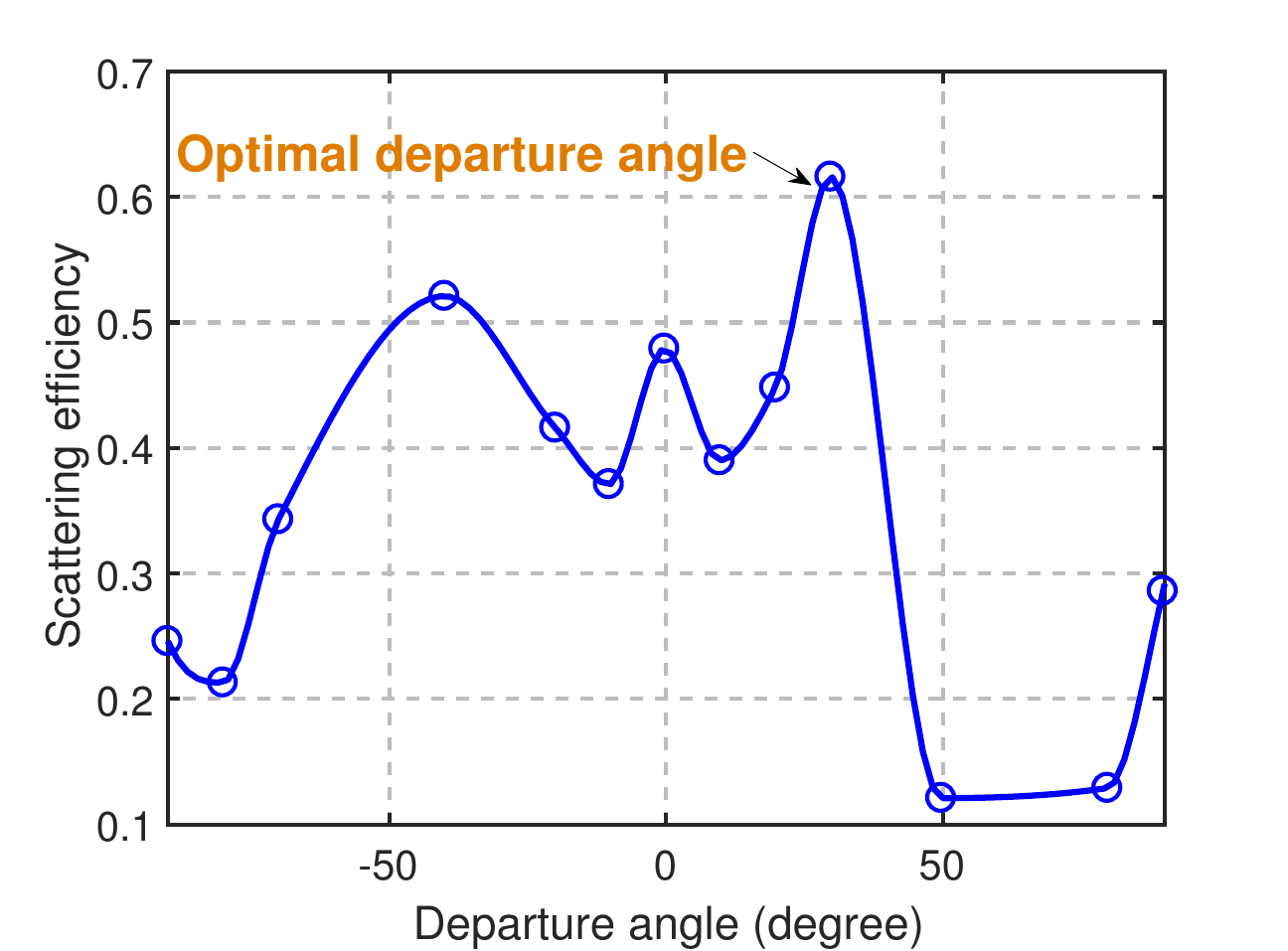}
			\vspace{-0.3cm}
			\label{eff_reflection_region}
	\end{minipage}}
	\subfigure[Rx 1 is in the refraction region]{
		\begin{minipage}[b]{0.45\textwidth}
			\centering
			\includegraphics[width=0.7\textwidth]{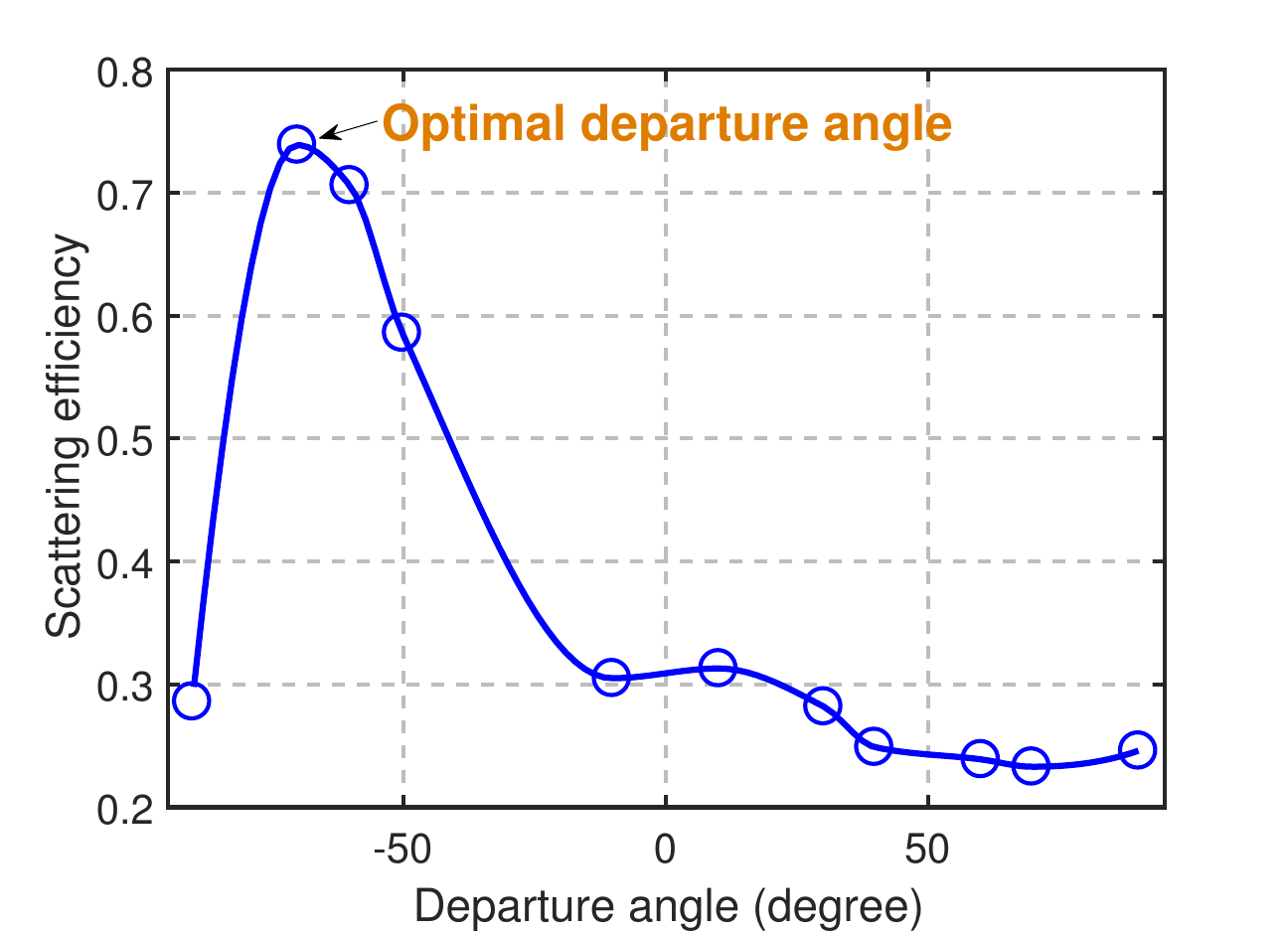}
			\vspace{-0.3cm}
			\label{eff_refraction_region}
	\end{minipage}}
	\vspace{-0.4cm}
	\caption{Scattering efficiency vs. the angle of departure from the IOS.}
	\label{eff}
	\vspace{-3mm}
\end{figure*}

\begin{figure*}[!t]
		\centering
		\includegraphics[width=0.4\textwidth]{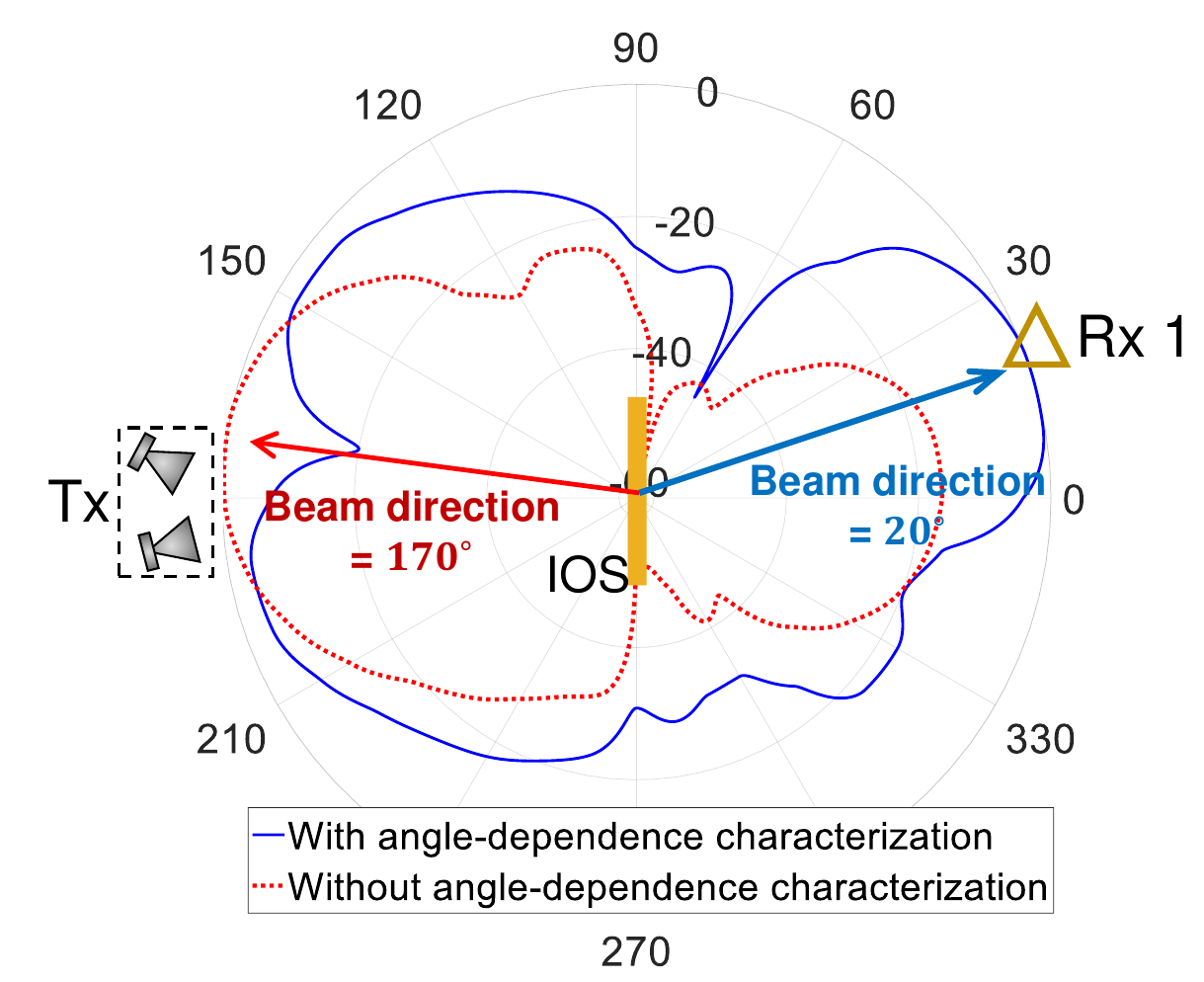}
		\vspace{-4mm}
		\caption{Comparison of the scattered beams when the angle-dependence response of the IOS is or is not considered.}
		\label{cmp_angle_dependence}
		\vspace{-7mm}
	\end{figure*}

\vspace{-2mm}
	 As mentioned, we analyze the beam pattern that corresponds to the receiver Rx 1. Since the size of the room is not large enough to measure the far-field beam pattern of the whole IOS, the experiments are conducted by considering a portion of the IOS (subarray), which consists of 8 rows and 5 columns of IOS elements, whose total size is $14.35\times 11.36 \times 0.71$~cm$^3$. To simplify the configuration of the IOS subarray, it is split into $8$ groups. Each group comprises one row of IOS elements, and the IOS elements within one group are set to the same state. To avoid undesired scattering, in addition, the rest of the IOS is covered with wave absorbing material, as shown in Fig.~\ref{exp_setting}. Therefore, the circular measurement platform is centered in correspondence of the IOS subarray, as shown in Fig.~\ref{pattern_layout}. The locations and orientations of the transmit and receive antennas are illustrated in Fig.~\ref{pattern_layout} as well. In particular, the observation distance with respect to the center of the IOS subarray is $0.97$~m. This implies that the observation points are in the Fraunhofer far-field distance of the IOS subarray, which is approximately equal to $0.8$~m at the measurement frequency of $3.6$~GHz. The results illustrated in Fig.~\ref{beam_transmission}-\ref{cmp_angle_dependence} are all obtained based on the experimental layout shown in Fig.~\ref{pattern_layout}.
	

	The measured radiation patterns are illustrated in Fig.~\ref{beam_transmission} and Fig.~\ref{beam_reflection} when the IOS is configured to steer the beam towards the refraction and reflection regions, respectively. In the figures, we see that the IOS prototype offers a narrow HPBW of about $10^\circ$ and that the main lobe is directed towards the desired direction.
	

In Fig.~\ref{SLL}, we evaluate the SLL of the reradiated beam when the IOS steers the signal towards the intended receiver, which can be located in the reflection and refraction sides of the surface. We see that the SLL highly depends on the desired angle of departure, i.e., reradiation. The IOS prototype offers very good SLL performance when the angle of reradiation lies in $[-60^\circ,-40^\circ]$, for both reflection and refraction. The experimental characterization in Fig.~\ref{SLL} provides important information on the deployment and the association between IOSs and UEs. If, for example, several IOSs are deployed in the network (see, e.g.,~\cite{BMM_arxiv}), the UEs may be associated to the IOS that offers the lowest SLL while ensuring that correct steering of the main lobe.

In Fig.~\ref{eff}, we illustrate the scattering efficiency of the IOS prototype as a function of the angle of departure (in the reflection and refraction regions). We observe a significant dependency with the target angle of departure. In the reflection region, the efficiency is maximized for angles that lie in the range $[20^\circ,40^\circ]$. In the refraction region, the efficiency is maximized for angles that lie in the range $[-80^\circ,-60^\circ]$. This information can also be utilized to appropriately associate the UEs that are distributed throughout the network with the available IOSs.
	
	\begin{table}[!tpb]
		\small
		\centering
		\caption{\normalsize{Simulation parameters for Fig.~\ref{cmp_angle_dependence}}}
		\label{sim_par_beam}
		{
			\begin{tabular}{|l|l|}	
				\hline	
				\textbf{Parameters} & \textbf{Values} \\
				\hline\hline
				Part number of Tx antennas & LB-880\\
				\hline
				Part number of the Rx antenna & Cylindrical antenna with a gain of $3$~dBi\\
				\hline
				Size of IOS & $8$ rows and $5$ columns with $40$ elements \\
				\hline
				Structure of IOS elements & As shown in Section II-C \\
				\hline
				Grouping strategy of the IOS & $8$ groups, each with one row of IOS elements \\
				\hline
				Deployments of Tx antennas, Rx antenna 1, and the IOS & As shown in Fig. 9 \\
				\hline
		\end{tabular}}
	\end{table}
	
	To further demonstrate the importance of characterizing the impact of the angle of incidence on the steering capabilities of the IOS, we compare the beam patterns by taking or not taking into account the angle-dependent behavior of the reflection and transmission coefficients in (\ref{cons_state}). {The simulation parameters are summarized in Table~\ref{sim_par_beam}.} The results are illustrated in Fig.~\ref{cmp_angle_dependence}. Specifically, the curve that accounts for the impact of the angle of incidence is obtained by solving the optimization problem in (\ref{h_bf}) when imposing the angle-dependent constraint in (\ref{cons_state}). The curve that does not account for the impact of the angle of incidence is, on the other hand, obtained by replacing the constraint in (\ref{cons_state}) with $(\Gamma_r^{m,k},\Gamma_t^{m,k})\in\left\{\left(\Gamma_r(0,0,s_m),\Gamma_t(0,0,s_m)\right)\right\}_{s_m\in \mathcal{S}}, \forall m$. In other words, as a case study, the reflection and transmission coefficients that correspond to normal incidence are utilized. Figure~\ref{cmp_angle_dependence} shows that the obtained radiation patterns are different, and that the directions of the corresponding main lobes (i.e., the beam directions) are different from each other as well. This result proves the importance of utilizing appropriate EM models for optimizing IOSs.

	\vspace{-0.4cm}
	\subsection{Analysis of the Data Rate}
	\label{exp_data_rate}

	\begin{figure*}[!tpb]
		\centering
		\includegraphics[width=0.25\textwidth]{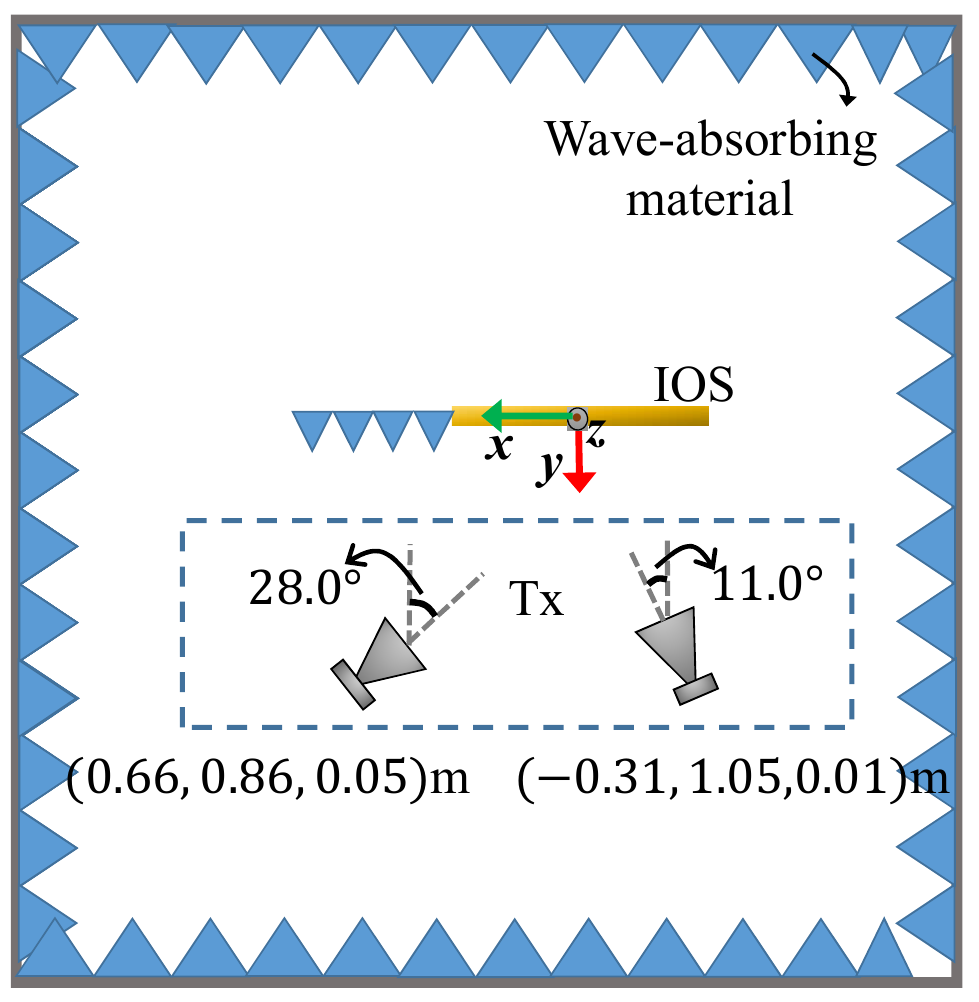}
		\vspace{-0.2cm}
		\caption{Experimental layout for measuring the data rate. The figure shows the position of the two transmit antennas.}
		\label{layout_data}
		\vspace{-9mm}
	\end{figure*}

\begin{figure*}[!tpb]
	\centering
	\subfigure[]{
		\begin{minipage}[b]{0.3\textwidth}
			\centering
			\includegraphics[width=1\textwidth]{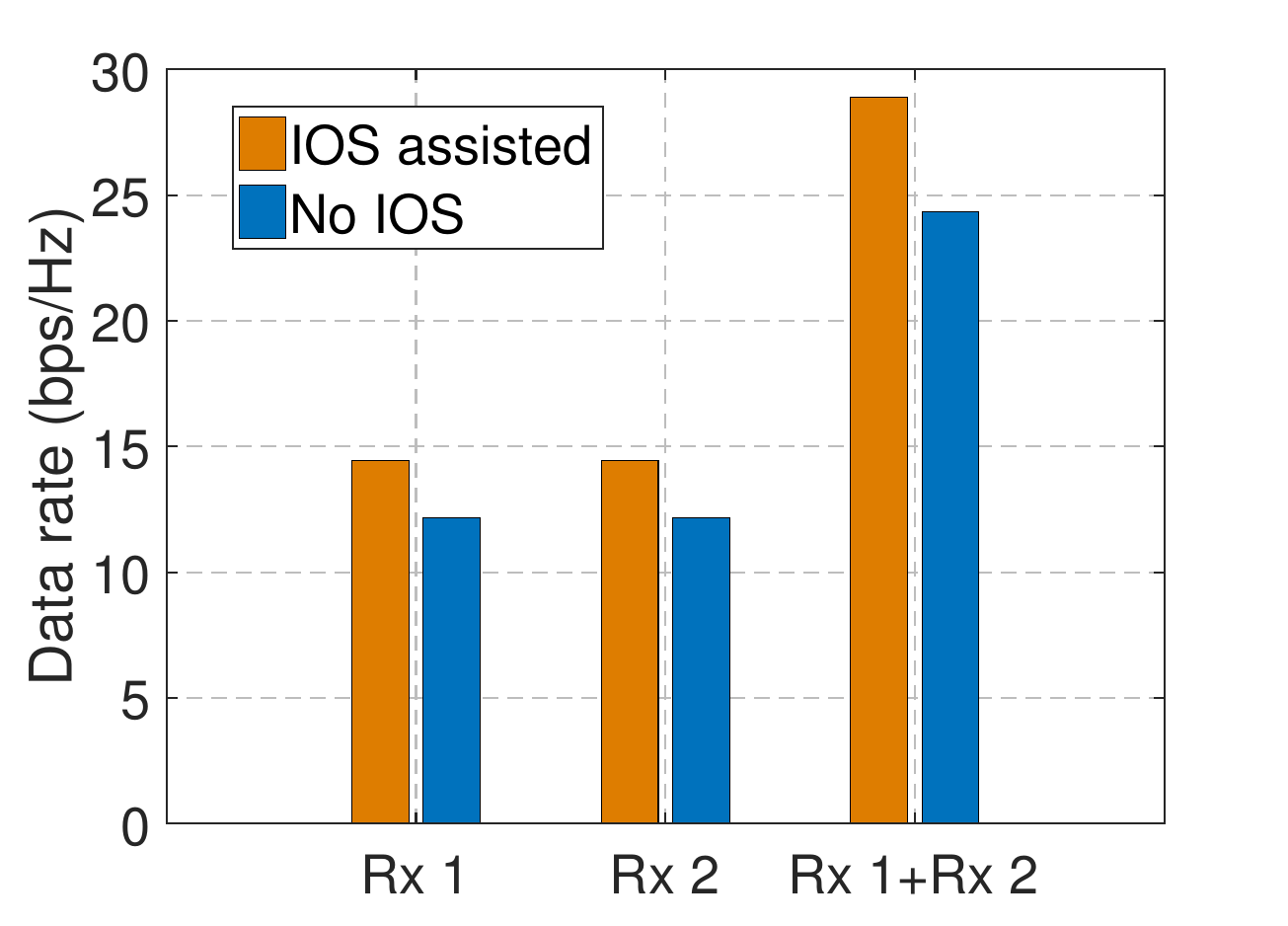}
					\vspace{-0.45cm}
			\vspace{-0.6cm}
			\label{data_rate_rr}
	\end{minipage}}
	\subfigure[]{
		\begin{minipage}[b]{0.3\textwidth}
			\centering
			\includegraphics[width=1\textwidth]{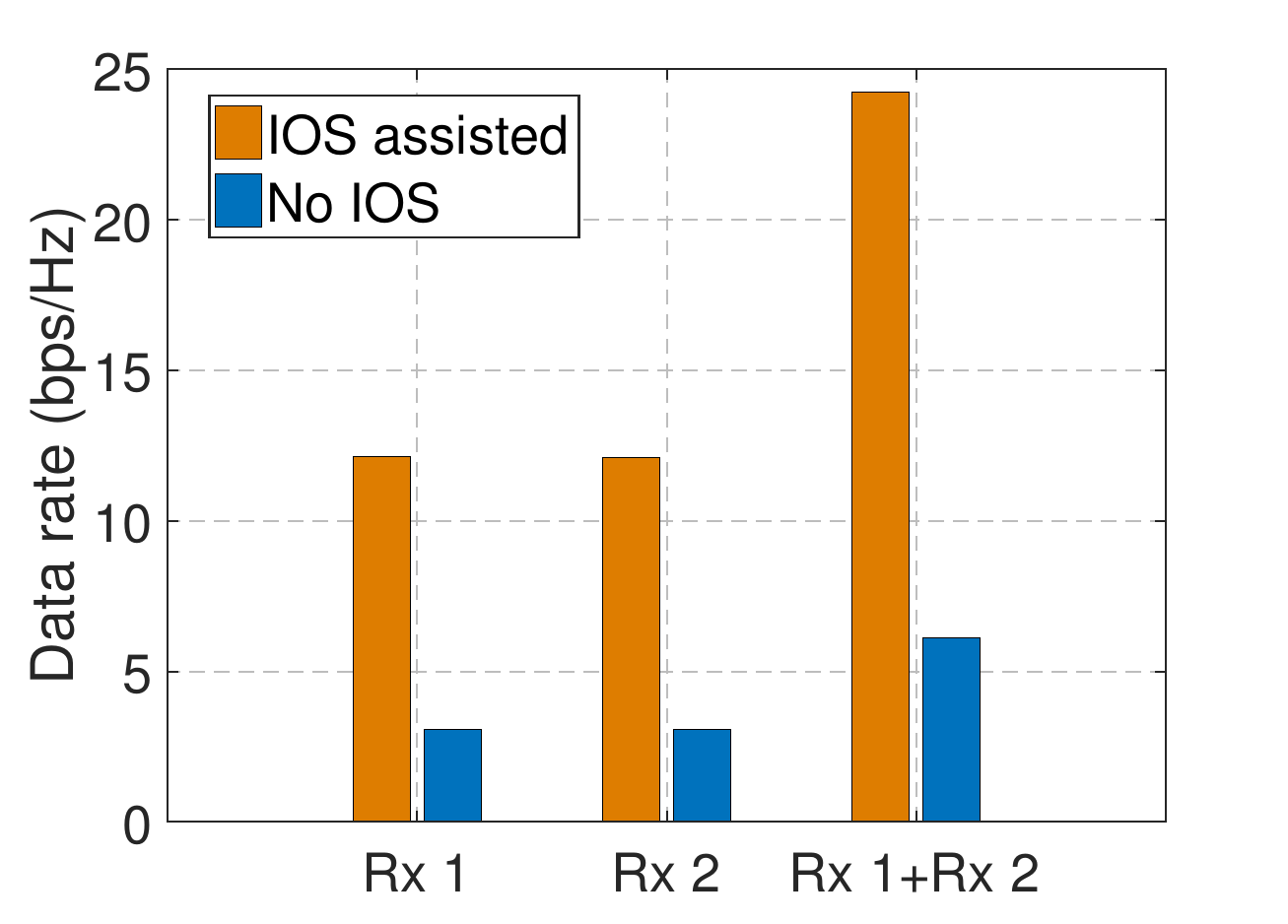}
					\vspace{-0.45cm}
			\vspace{-0.6cm}
			\label{data_rate_tt}
	\end{minipage}}		
	\subfigure[]{
		\begin{minipage}[b]{0.3\textwidth}
			\centering
			\includegraphics[width=1\textwidth]{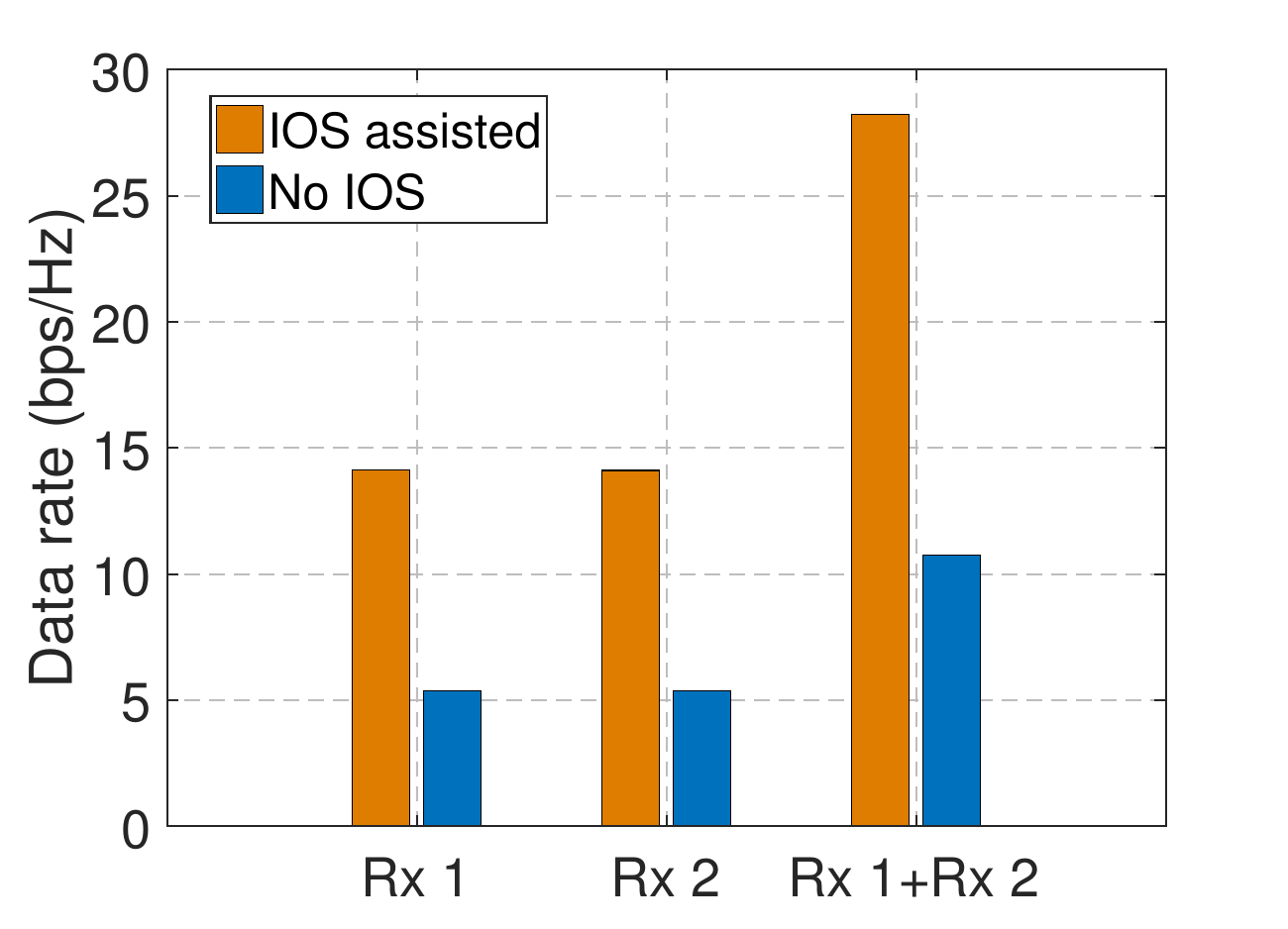}
					\vspace{-0.45cm}
			\vspace{-0.6cm}
			\label{data_rate_rt}
	\end{minipage}}
	\vspace{-0.4cm}
	\caption{Data rate of the prototype, where (a) both Rxs are in the reflection region, (b) both Rxs are in the refraction region, and (c) one Rx is located in the reflection region and one Rx is located in the refraction region.}
	\vspace{-3mm}
	\label{data_rate_fig}
\end{figure*}

\begin{figure*}[!t]
	\centering
	\includegraphics[width=0.45\textwidth]{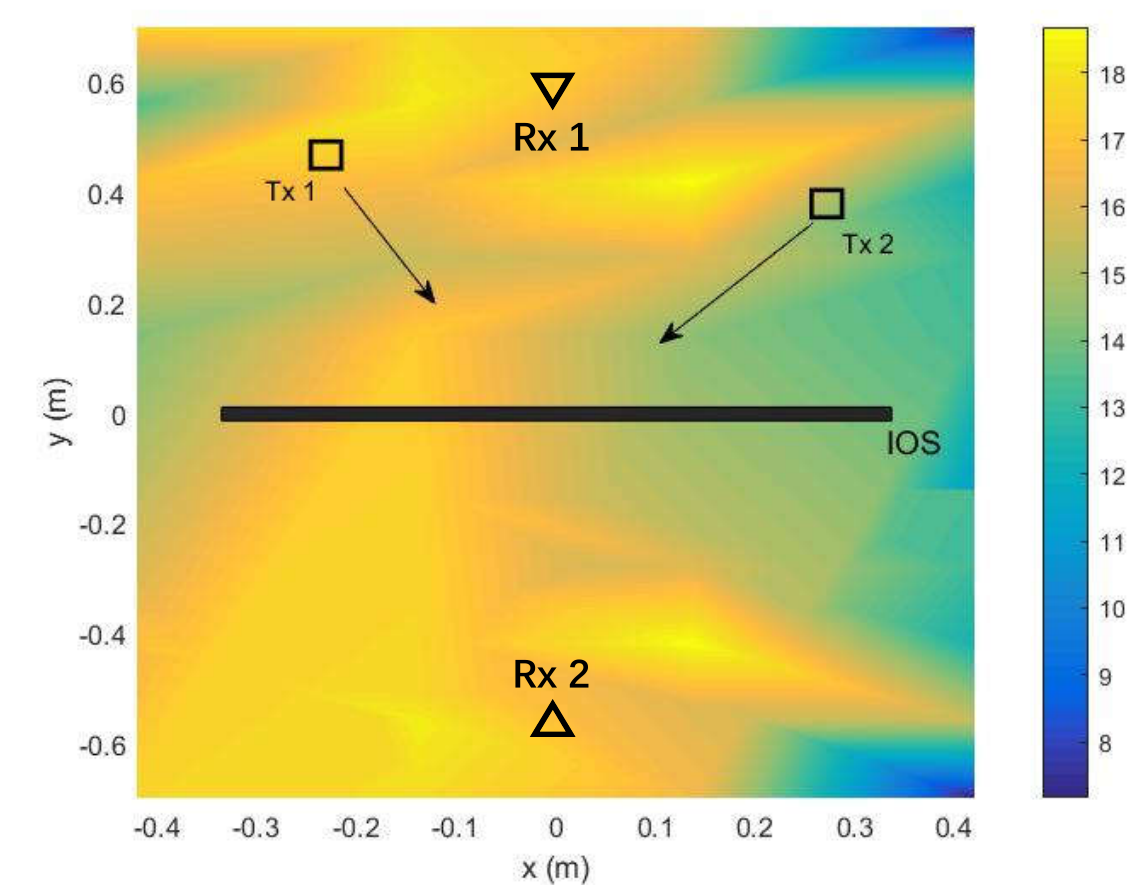}
	\vspace{-0.3cm}
	\caption{Data rate distribution in the reflection and refraction sides of the IOS.}
	\label{capacity_distribution}
	\vspace{-5mm}
\end{figure*}

\begin{figure*}[!t]
		\centering
		\includegraphics[width=0.5\textwidth]{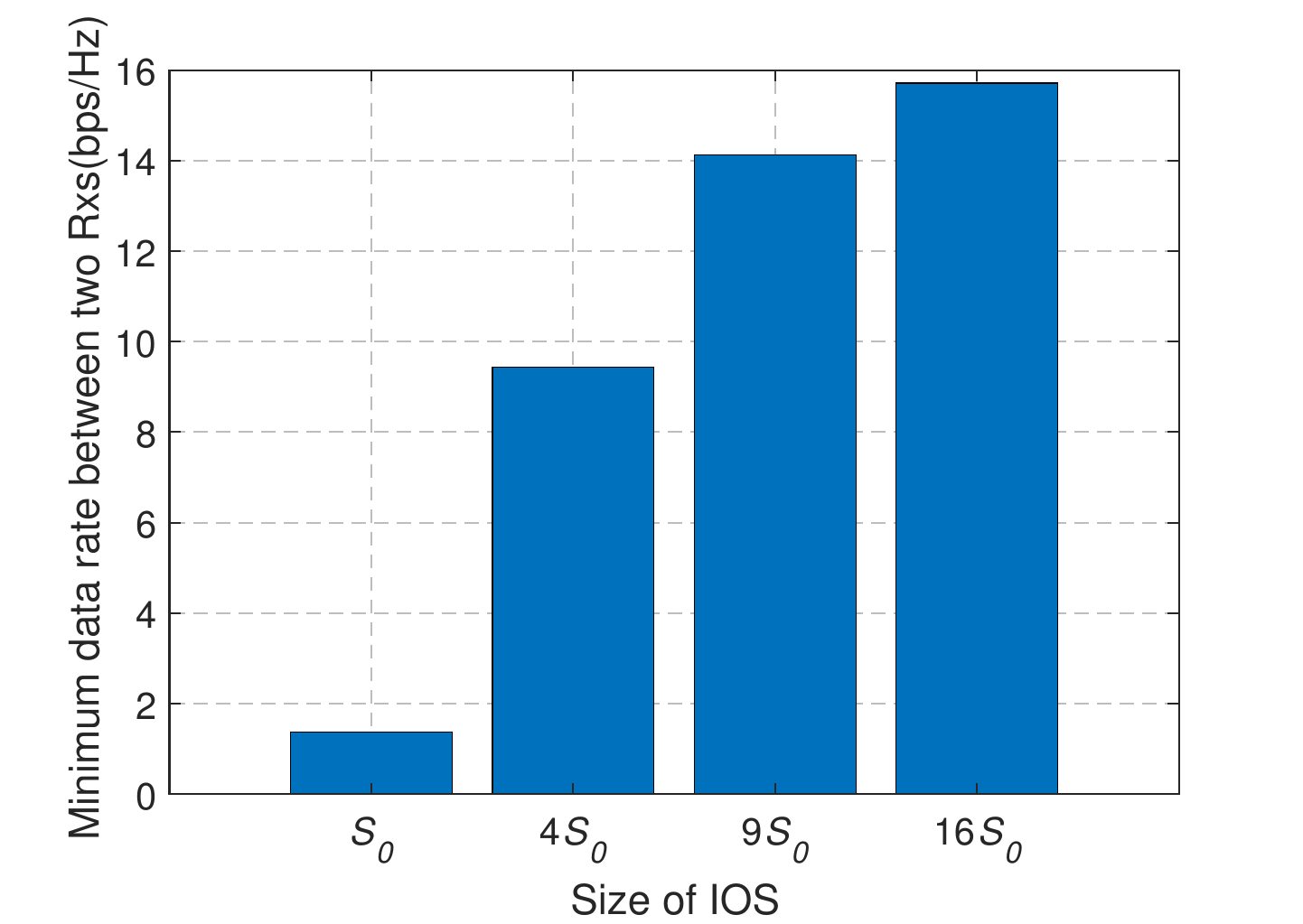}
		\vspace{-0.2cm}
		\caption{Minimum data rate of two Rxs vs. the size of IOS, with $S_0=14.35\times 11.36$~cm$^2$.}
		\label{data_rate_vs_size}
		\vspace{-8mm}
	\end{figure*}

{In this section, we analyze the data rate of the implemented IOS-aided multi-user system.} For this study, we consider the \textit{entire} IOS prototype, which is deployed within a room whose walls are made of aluminum, in order reduce the impact of external EM interference, and are coated with wave absorbing material, in order to avoid undesired reflections. The experimental layout, and the positions and orientations of the transmit antennas with respect to the IOS are illustrated in Fig.~\ref{layout_data}. To simplify the control circuit, the elements of the IOS are split into 16 groups. Each group comprises $5 \times 8$ IOS elements, which are all set to the same state. {The SINR threshold $\gamma_{th}$ is set to $6$~dB, which is in agreement with other research works~\cite{MMCA_2022}.} To analyze the full-dimensional communication capabilities of the IOS, three different deployments for the receivers are considered: 1) Two receivers are placed in the reflection region of the IOS at the locations $(0.22,1.17,-0.24)$~m and $(1.06,0.49,-0.24)$~m, 2) two receivers are in the refraction region at the locations $(0.22,-0.97,-0.24)$~m and $(1.06,-0.49,-0.24)$~m, and 3) one receiver is located in the reflection region of the IOS and one receiver is located in the refraction region at the locations $(0.22,1.17,-0.24)$~m and $(1.06,-0.49,-0.24)$~m, respectively. Since the Fraunhofer far-field distance of the entire IOS is approximately equal to $15.36$~m at $3.6$~GHz, the three considered deployments correspond to transmitters and receivers located in the radiative near-field region of the IOS. {The experimental data rate obtained with the prototype is illustrated in Figs.~\ref{data_rate_fig} and Fig.~\ref{data_rate_vs_size}, and the simulated data rate obtained with the aid of simulations is displayed in Fig.~\ref{capacity_distribution}.} In the considered scenario, specifically, the objective function in (\ref{obj}) is the minimum of the data rate of the two receivers.

\begin{table}[!tpb]
	\small
	\centering
	\caption{\normalsize{ {Simulation parameters for Fig.~\ref{capacity_distribution}}}}
	\label{sim_par_rate}
	{
		\begin{tabular}{|l|p{7cm}|}	
			\hline	
			\textbf{Parameters} & \textbf{Values} \\
			\hline\hline
			Part number of Tx antennas & LB-880\\
			\hline
			Part number of the Rx antenna & Cylindrical antenna with a gain of $3$~dBi\\
			\hline
			Size of IOS & $32$ rows and $20$ columns with $640$ elements \\
			\hline
			Structure of IOS elements & As shown in Section II-C \\
			\hline
			Grouping strategy& $16$ groups of IOS elements, each with $8$ rows and $5$ columns\\
			\hline
			Deployments of Tx antennas and the IOS & As shown in Fig. 15 \\
			\hline
			Noise power & $-96$~dBm\\
			\hline 
			Maximum transmit power for each antenna& $200$~mW\\
			\hline
			Pathloss exponent & 2\\
			\hline
			Carrier frequency & $3.6$~GHz\\
			\hline 
	\end{tabular}}
\end{table}
	
	From Fig.~\ref{data_rate_fig}, we evince that the data rate can be significantly increased in the presence of an IOS. We see that the two receivers, regardless of their location with respect to the IOS, can achieve a similar data rate. This is because the minimum of the data rate of the two receivers is considered as the utility function. The setup in the absence of IOS (i.e., ``No IOS") is obtained by covering the IOS with wave absorbing material. In agreement with Figs.~\ref{response_IOS_ele} and~\ref{incident_angle}, the receivers in the refraction region obtain a more pronounced improvement of the data rate with respect to the ``No IOS" case, thanks to the stronger transmission coefficient.

	In Fig.~\ref{capacity_distribution}, we report simulation results that illustrate the data rate distribution, in the radiative near-field region of the IOS across an area of approximate size equal to $1.2 \times 0.8$ square meters. The two receivers are located in the positions $(x,y,-0.24)$ m and $(x,-y,-0.24)$ m, where $(x,y)$ lie in the region depicted in Fig.~\ref{capacity_distribution}. {Other simulation parameters are given in Table~\ref{sim_par_rate}.} The data rate is calculated at every point $(x,y)$ and is illustrated as a color map. We see that a good data rate is obtained around the two receivers.
	  
	{In Fig.~\ref{data_rate_vs_size}, we record the minimum data rate between the two Rxs as a function of the size of the IOS. The transmitters are deployed as illustrated in Fig.~\ref{layout_data}. The two receivers are located in the positions $(1.06,0.49,-0.24)$~m and $(0.11,0.97,-0.24)$~m, respectively. From Fig.~\ref{data_rate_vs_size}, we find that the data rate is positively correlated with the size of the IOS. This is because a larger IOS can capture more transmitted power.}
	
	\begin{figure}[!ht]
		\centering
		\includegraphics[width=0.5\textwidth]{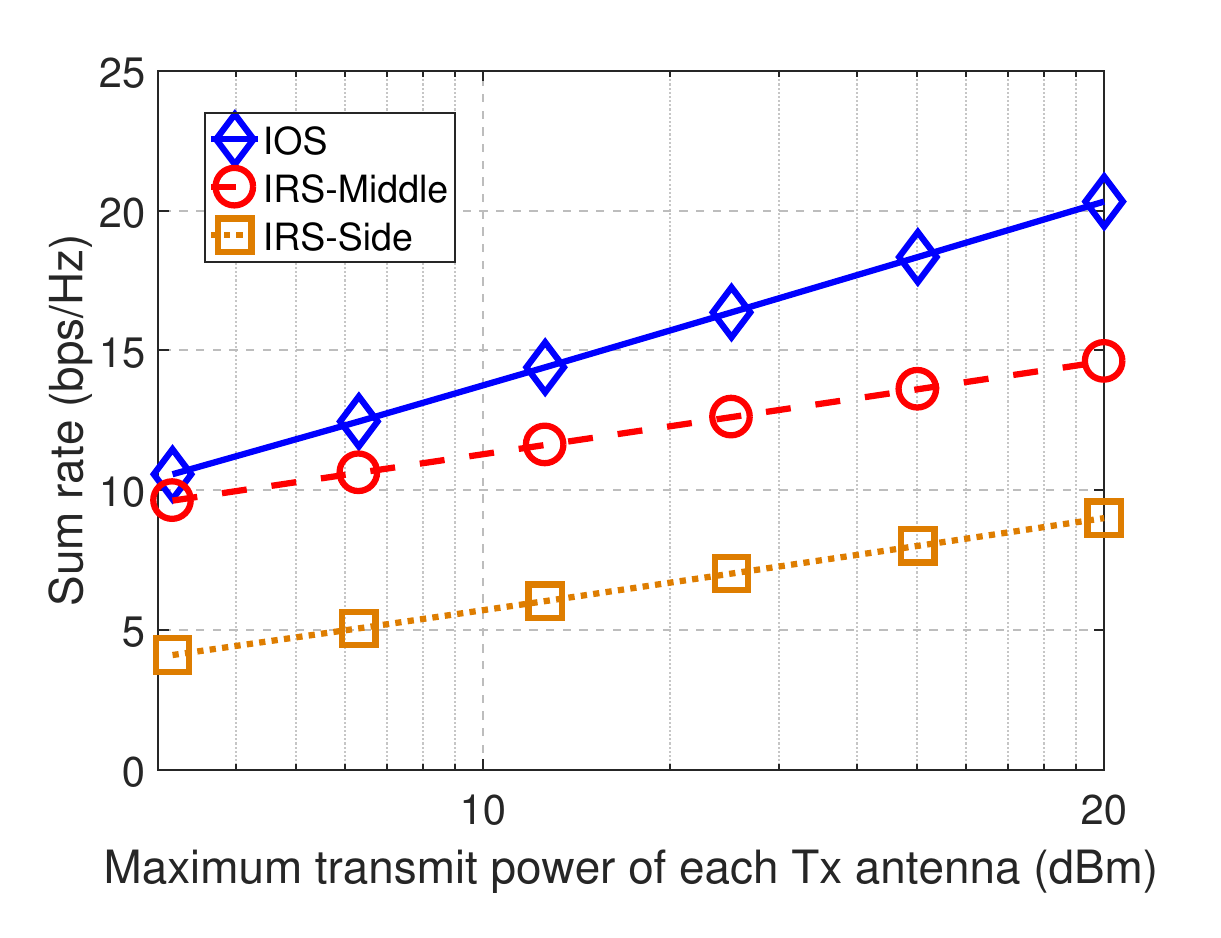}
		\vspace{-2mm}
		\caption{Comparison of sum rate between IOS and IRS.}
		\label{3cmp}
	\end{figure}
	
	\begin{table}[!htpb]
		\small
		\centering
		\caption{\normalsize{Simulation Parameters}}
		\label{sim_par_cmp}
		{
		\begin{tabular}{|l|p{7cm}|}	
			\hline	
			\textbf{Parameters} & \textbf{Values} \\
			\hline\hline
			Part number of Tx antennas & LB-880\\
			\hline
			Part number of the Rx antenna & Omni-directional with a gain of $3$~dBi\\
			\hline
			Size of IOS and IRS & $32$ rows and $20$ columns with $640$ elements \\
			\hline
			Structure of IOS elements & As shown in Section II-C \\
			\hline
			Size and radiation efficiency of IRS elements & The same as the IOS elements \\		
			\hline
			Grouping strategy& $16$ groups of metasurface elements, each with $8$ rows and $5$ columns\\
			\hline
			Deployments of Tx antennas & Tx antenna 1: $(0.66,0.86,2)$~m, Tx antenna 2: $(-0.31,1.05,2)$~m\\
			\hline
			Deployments of Rxs & Rx 1: $(0.4,1.2,-0.24)$~m, Rx 2: $(-0.6,-1,-0.24)$~m\\
			\hline
			Deployment of IOS & On the $xoz$ plane with its center coincides with the origin\\
			\hline
			Deployment of IRS in IRS-Middle scheme& On the $xoz$ plane with its center coincides with the origin\\
			\hline
			Deployment of IRS in IRS-Side scheme& Parallel to the $yoz$ plane, and the coordinate of its center is $(-1,0,0)$~m\\
			\hline
			Noise power & $-90$~dBm\\
			\hline
			Carrier frequency & $3.6$~GHz\\
			\hline
		\end{tabular}}
	\end{table}

	{In Fig.~\ref{3cmp}, we compare the sum rate achieved by the IOS against the sum rate of the reflective-type RIS~(IRS). To demonstrate the advantage of the IOS, two deployments for the IRS are considered: (1) \emph{IRS-Middle}: the location of the IRS is the same as that of the IOS, and the Rxs are located on two sides of the surface (2) \emph{IRS-Side}: the IRS is deployed such that all the Rxs and the Tx are located on the same side of the BS. For all the schemes, we assume that the direct links between the Tx and the Rxs are blocked. The other simulation parameters are listed in Table~\ref{sim_par_cmp}. As shown in Fig.~\ref{3cmp}, we see, for all the considered schemes, that the sum rate is positively correlated with the transmit power. Besides, it is observed that, compared with the IRS-Middle scheme, the IOS achieves a higher sum rate, since the Rxs on both sides of the surface can be served. In addition, the IOS also outperforms the IRS-Side scheme, indicating that the IOS can be appropriately optimized to outperform the IRS.}

	\vspace{-0.3cm}
	\section{Conclusion}
	\label{conc}
	\vspace{-0.2cm}
	In this paper, we have proposed a new reflection-refraction model for the reconfigurable elements of IOSs. The circuit-based model has been verified through full-wave EM simulations. Based on the obtained equivalent circuit model, we have demonstrated the influence of the angle of incidence and the IOS structure on the reflection and transmission coefficients. {Also, we have studied the asymmetry between the reflection and transmission coefficients.} Moreover, the proposed circuit-based model has been integrated into a mathematical framework for jointly optimizing the digital beamformer at the transmitter and the analog beamformer at the IOS. To verify the theoretical findings and evaluate the performance of the full-dimensional beamforming, we have implemented an IOS hardware prototype and have deployed an IOS-assisted wireless communication testbed, and have experimentally characterized the beam pattern of the IOS and the achievable rate. Based on the proposed equivalent circuit model and the experiments conducted, the following main conclusions are obtained:
	
	\begin{itemize}
		\item {The angle of incidence has a non-negligible impact on the reflection and transmission coefficients, which should be taken into account in the hybrid beamforming design for improving the system performance.}
		
		\item In network deployments in which multiple IOSs and receivers are available, the receivers should communicate with the aid of the IOS that corresponds to a moderate angle of reradiation.  {This configuration helps reduce the SLL and helps improve the scattering efficiency, which in turn is beneficial for full-dimensional communications.}
		
		\item The experimental results show that the IOS can steer the reflected and refracted far-field beams over $360^\circ$. Also, ubiquitous coverage can be obtained in the radiative near-field region as well. 
		
	\end{itemize} 

\begin{appendices}
	\vspace{-0.5cm}

	\begin{figure*}[!t]
		\centering
		\includegraphics[width=0.35\textwidth]{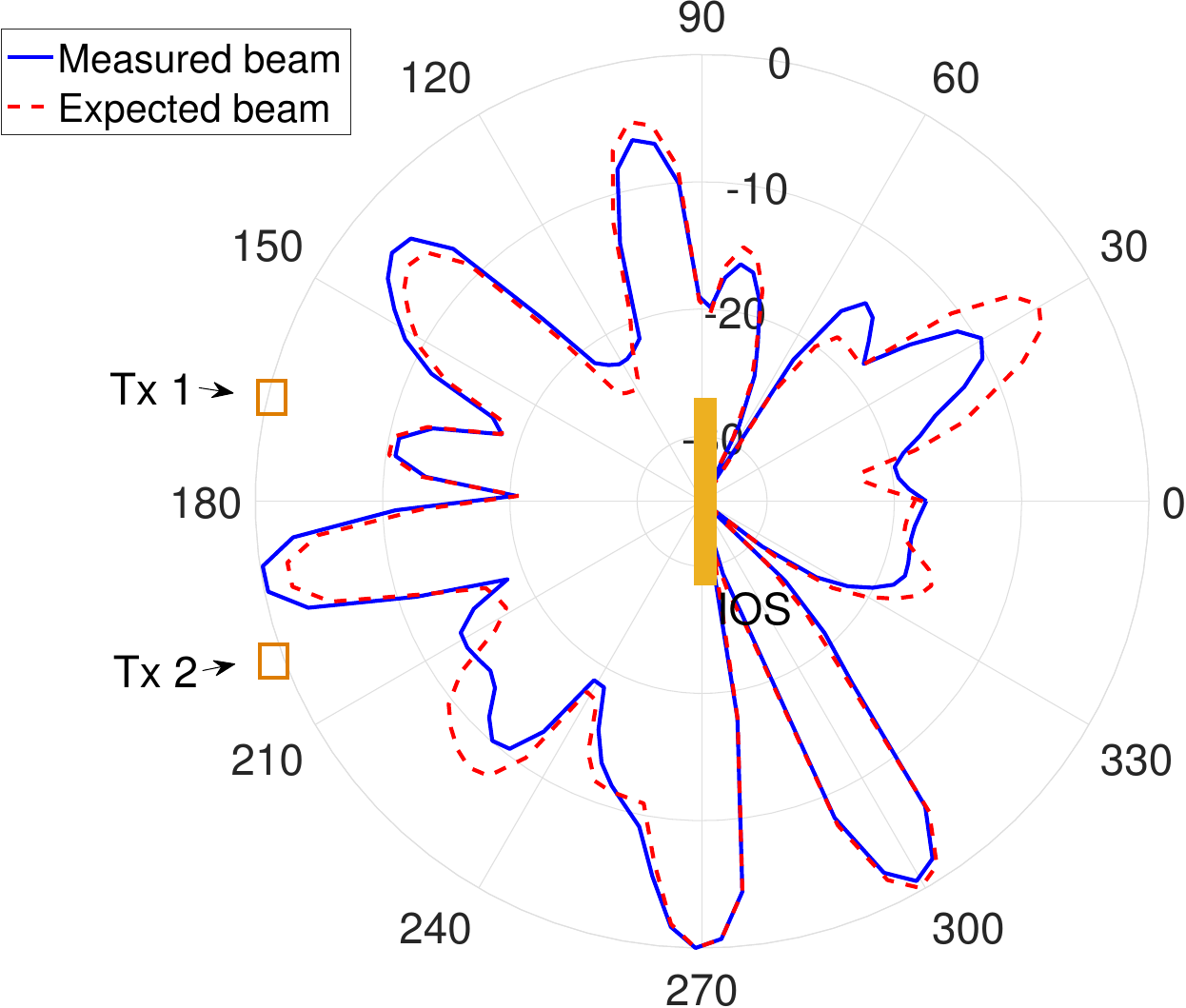}
		\vspace{-4mm}
		\caption{Comparison between measured and expected scattered beams.}
		\label{verify_pattern}
				\vspace{-9mm}
	\end{figure*}
	\section{Verification of the model for the scattered beam in (\ref{pattern})}
	\label{app_model_beam}
	
	To demonstrate the validity of the model for the scattered beam in (\ref{pattern}), we select a specific IOS configuration and a given digital beamformer, and measure the corresponding scattered beam. In Fig.~\ref{verify_pattern}, we compare the measured scattered beam and the expected scattered beam based on (\ref{pattern}). We see that the measured beam matches well with the expected beam, which indicates that the scattered beam can be predicted with the proposed model in (\ref{pattern}). 

\end{appendices}	

\end{document}